\documentclass[10pt,journal,final,twocolumn,]{IEEEtran}
\IEEEoverridecommandlockouts

\usepackage{amsmath,amssymb,amsfonts}
\usepackage{cite}
\usepackage{graphicx}
\usepackage{epstopdf}

\usepackage{booktabs}
\usepackage{graphicx}
\usepackage{textcomp}
\usepackage{xcolor}
\usepackage{times}
\usepackage{subfigure}         
\usepackage{amssymb,amsmath}
\usepackage{acronym}  
\usepackage{balance}

\usepackage{bm}         
\usepackage{algorithm} 
\usepackage{algorithmic}

\usepackage{lipsum}
\usepackage{stfloats}

\usepackage{url}

\usepackage{amsmath}
\usepackage{mathrsfs}

\newtheorem{proposition}{\bf Proposition}

\newtheorem{remark}{\bf Remark}

\acrodef{ofdm}[OFDM]{orthogonal frequency division multiplexing}%
\acrodef{miso-ofdm}[MISO-OFDM]{multi-input single-output orthogonal frequency division multiplexing}%

\acrodef{ris}[RIS]{reconfigurable intelligent surface}%
\acrodef{qos}[QoS]{quality of service}%
\acrodef{idft}[IDFT]{inverse discrete Fourier transform}%
\acrodef{dft}[DFT]{discrete Fourier transform}%
\acrodef{cp}[CP]{cyclic prefix}%
\acrodef{csi}[CSI]{channel state information}%
\acrodef{awgn}[AWGN]{additive white Gaussion noise}%

\acrodef{qcqp}[QCQP]{quadratically constrained quadratic program}%
\acrodef{qp}[QP]{quadratic program}%
\acrodef{bs}[BS]{base station}%
\acrodef{ap}[BS]{base station}%
\acrodef{aps}[APs]{access points}%
\acrodef{qos}[QoS]{quality of service}%
\acrodef{ue}[UE]{user equipment}%
\acrodef{snr}[SNR]{signal-to-noise ratio}%
\acrodef{mmwave}[mmWave]{millimeter-wave}%
\acrodef{snr}[SNR]{signal-to-noise ratio}%

\acrodef{sinr}[SINR]{signal-to-interference-plus-noise ratio}%

\acrodef{ser}[SER]{symbol error rate}%
\acrodef{rc}[RC]{reflection coefficient}%
\acrodef{uavs}[UAVs]{unmanned aerial vehicles}%
\acrodef{mimo}[MIMO]{multiple-input multiple-output}%
\acrodef{noma}[NOMA]{non-orthogonal multiple access}%

\acrodef{ace}[ACE]{adaptive cross-entropy}%
\acrodef{wsr}[WSR]{weighted sum-rate}%
\acrodef{udn}[UDN]{ultra-dense network}%
\acrodef{Udn}[UDN]{Ultra-dense network}%

\def\BibTeX{{\rm B\kern-.05em{\sc i\kern-.025em b}\kern-.08em
    T\kern-.1667em\lower.7ex\hbox{E}\kern-.125emX}}

\begin{document}
\title{Distance-Aware Precoding for Near-Field \\Capacity Improvement}
\author{\IEEEauthorblockN{Zidong Wu, \emph{Student Member, IEEE}, Mingyao Cui, \emph{Student Member, IEEE}, {Zijian Zhang}, \emph{Student Member, IEEE}, Linglong Dai, \emph{Fellow, IEEE}}

\thanks{All authors are with the Beijing National Research Center for Information Science and Technology (BNRist) as well as the Department of Electronic Engineering, Tsinghua University, Beijing 100084, China (e-mails: \{wuzd19, cmy20, zhangzj20\}@mails.tsinghua.edu.cn, daill@tsinghua.edu.cn).}
}

\maketitle
\begin{abstract}
Extremely large-scale MIMO (XL-MIMO) is a promising technology to improve the capacity for future 6G networks. With a very large number of antennas, the near-field property of XL-MIMO systems becomes dominant. Unlike the classical far-field line-of-sight (LoS) channel with only one available data stream, the significantly increased degrees of freedom (DoFs) are available in the near-field LoS channel. However, limited by the small number of radio frequency (RF) chains, the existing hybrid precoding architecture widely used for 5G is not able to fully exploit the extra DoFs in the near-field region. In this paper, the available DoFs and the capacity of the near-field LoS channel are theoretically analyzed at first. Then, to exploit the near-field effect as a new possibility for capacity improvement, the distance-aware precoding (DAP) scheme is proposed. We develop the DAP architecture, where a dedicated selection circuit is inserted to connect phase shifters and RF chains. Moreover, each RF chain can be flexibly configured to active or inactive according to the distance-related DoFs in the proposed DAP architecture. Based on the developed DAP architecture, a DAP algorithm is proposed to optimize the number of activated RF chains and precoding matrices to match the increased DoFs in the near-field region. Finally, simulation results verify that, the proposed DAP scheme can efficiently utilize the extra DoFs in the near-field region to improve the spectrum efficiency and the energy efficiency as well.

\end{abstract}
\section{Introduction}
Due to the explosively growing demand for communication capacity, spatial multiplexing is regarded as a key technology to significantly increase the spectrum efficiency \cite{sun'14'j}. By exploiting the spatial degrees of freedom (DoFs) with spatial multiplexing, massive multiple-input multiple-output (MIMO) is the key enabler to increase the capacity by orders of magnitude for 5G. To further exploit the promising potential of spatial DoFs, massive MIMO for 5G is evolving to extremely large-scale MIMO (XL-MIMO) for future 6G communications \cite{bjornson'2019'j}.
\par
The evolution from massive MIMO to XL-MIMO not only implies a significant increase in the number of antennas, but also leads to a fundamental change of the characteristics in the electromagnetic field. Specifically, since the number of antennas of massive MIMO systems is usually not very large, the classical \emph{far-field} MIMO channel model is widely adopted, which is based on planar wave assumptions \cite{Ayach'14'j}. By contrast, with the dramatically increasing number of antennas in XL-MIMO communication systems, receivers may lie in the \emph{near-field} region of the transmitter, which is defined by the area within the Rayleigh distance \cite{Sherman'62'j}. For example, if we consider an XL-MIMO system with 1000 linear antennas with half wavelength spacing, the Rayleigh distance reaches 500 m at 300 GHz, which covers a large area of a standard cell. Unlike the classical \emph{far-field} channel based on planar wave assumptions, the \emph{near-field} channel should be modeled based on spherical wave assumptions in XL-MIMO communications, leading to significant near-field effect \cite{cui'21}. Thus, traditional far-field transmission techniques may suffer from a severe performance loss due to the mismatch between \emph{far-field} assumptions and \emph{near-field} channel property.
\subsection{Prior Works}\label{sec:intro prior}
Most of existing works on near-field communications focus on overcoming the performance degradation caused by the near-field effect \cite{headland'18't,zhang'21',cui'21}. To be specific, the steering beams are only related to the user angles under the planar wave assumptions. By contrast, due to the property of spherical waves, the beams are related to both the user angle and distance \cite{headland'18't}. The initial work \cite{zhang'21'} has studied the near-field beamforming techniques by designing the angle- and distance-dependent beams for narrow-band communications. To achieve the near-field beamforming in wideband systems, by exploiting the varied spherical wavefronts at different frequencies, a phase-delay focusing method was proposed in \cite{cui'21}.
\par
To acquire high-gain beamforming, channel estimation is essential for XL-MIMO communications. Several works \cite{cui'22'j,Yu'20'j,Wei'21'j,han'20'j} have studied how to overcome the performance loss of channel estimation in the near-field region. In \cite{cui'22'j}, a near-field channel estimation method was proposed to capture the sparsity in the joint angle-distance (polar) domain caused by spherical waves. Moreover, a more realistic hybrid-field channel model characterizing both far-field and near-field paths was investigated in \cite{Wei'21'j}, and the corresponding channel estimation method was also proposed. Existing works on near-field communications indicate that the near-field effect only induces performance degradation for wireless communications which is caused by the mismatch of the near-field channel model and classical far-field-based beamforming or channel estimation methods in the physical layer.
\par
Nevertheless, the recently developed electromagnetic information theory (EIT) indicates that, wireless communications can also benefit from the near-field effect \cite{zhangzz'21'}. Specifically, for far-field channels, due to the high attenuation of millimeter-wave and terahertz bands, the rank-one line-of-sight (LoS) path becomes predominant in the XL-MIMO channels, so only very limited spatial multiplexing gain can be achieved \cite{Davide'20'jsac}. Compared with the far-field LoS path with \emph{planar} wavefronts determined by a single spatial angle, the near-field LoS path with \emph{spherical} wavefronts contains a range of angles, which brings significantly increased spatial DoFs (or transmission modes) \cite{miller'00'j, Decarli'21}. Owing to the increased spatial DoFs, it is expected that the spectrum efficiency can be naturally enhanced in the near-field region.
\par 
Unfortunately, the widely considered hybrid precoding architecture for XL-MIMO can hardly benefit from the increased DoFs provided by the near-field LoS path. In the far-field MIMO channel, due to the limited DoFs, the reduced number of RF chains for the classical hybrid precoding is still larger than the spatial DoFs. Thus, the classical hybrid precoding with a much reduced number of RF chains can fully utilize the spatial DoFs to achieve the near-optimal spectrum efficiency in the far-field region \cite{Heath'16'j}. By contrast, this reduced number of RF chains is much smaller than the increased DoFs in the near-field region, resulting in a very limited number of data streams. That is to say, the classical hybrid precoding architecture designed for far-field communications can hardly utilize the increased spatial DoFs in the near-field region. To the best of our knowledge, this important problem of how to exploit the extra spatial DoFs brought by the near-field effect to improve the capacity has not been studied in the literature.
\subsection{Our Contributions}\label{sec:intro contribution}
To fill in this gap, in this paper, the theoretical capacity of the near-field channel is analyzed, and the precoding architecture for near-field MIMO communications is proposed\footnote{Simulation codes are provided to reproduce the results presented in this paper: http://oa.ee.tsinghua.edu.cn/dailinglong/publications/publications.html.}. The main contributions of this paper can be summarized as follows:

\begin{itemize}
\item We introduce the eigenproblem in the electromagnetic theory to investigate the communication capacity of the near-field LoS channel. First, the increased DoFs are estimated based on the eigenproblem of prolate spheroidal wave functions. The effectiveness of the approximation of the singular values of the near-field LoS channel is verified by numerical simulations. Then, the channel capacity of the near-field LoS channel is analyzed, which reveals the channel capacity improvement in the near-field region.
\item We propose a distance-aware precoding (DAP) architecture, which provides a new possibility to improve the capacity by utilizing the distance-related DoFs in the near-field region. The key idea is to adaptively adjust the number of RF chains to match the DoFs in the near-field region. Compared with the classical hybrid precoding architecture where the number of RF chains is fixed and limited, extra RF chains are equipped in the proposed DAP architecture, where each RF chain can be flexibly configured to active or inactive. To realize this structure, a dedicated selection circuit is inserted to connect phase shifters to activated RF chains, which controls the selection pattern between the RF chains and array elements.
\item Based on the DAP architecture, a DAP algorithm is proposed to improve the spectrum efficiency. In the proposed DAP algorithm, the number of activated RF chains is adaptively optimized according to the DoFs of the channel. Since the eigenvalues are close to each other in the near-field region, a near-field subarray partitioning algorithm is performed to optimize the selection matrix which represents the linking pattern between the RF chains and phase shifters to get a balanced subarray structure. Then, the corresponding digital precoder and analog precoder can be obtained utilizing the sub-connected feature. Finally, simulation results verify that, the proposed DAP scheme can efficiently utilize the extra DoFs to improve the spectrum efficiency and energy efficiency in the near-field region.
\end{itemize}
\subsection{Organization and Notation}\label{sec:intro org}
The remainder of the paper is organized as follows. Section \ref{sec:sys} introduces the XL-MIMO communication system model in the near-field region, and compares the relationship between the near-field LoS channel and far-field LoS channel. Section \ref{sec:theo} theoretically investigates the DoFs of the near-field LoS channel and provides an approximation method for the channel capacity. In Section \ref{sec:architecture}, the DAP architecture is proposed. In Section \ref{sec:Alg}, the DAP algorithm is illustrated in detail. Simulation results are provided in Section \ref{sec:simulation}, and conclusions are drawn in Section \ref{sec:conclusion}.  

\textit{Notations}: $\mathbb{C}$ denotes the set of complex numbers; ${[\cdot]^{-1}}$, ${[\cdot]^{T}}$, ${[\cdot]^{H}}$ and ${\rm diag}(\cdot)$ denote the inverse, transpose, conjugate-transpose and diagonal operations, respectively; $\|\cdot\|_F $ denotes the Frobenius norm of the matrix; $\angle[\cdot]$ denotes the angle of its complex argument; $\mathbf{I}_{N}$ is an $N\times N$ identity matrix; $[a]^+$ represents $\max \{0, a\}$; finally $\mathbf{1}_{L}$ indicates an $L$-length vector with all elements being 1.

\section{System Model}\label{sec:sys}

In this paper, we consider a single-user XL-MIMO communication system. The transmitter and receiver are equipped with $N_{\rm{t}}$ and $N_{\rm{r}}$ antennas, respectively. To enable the transmission of $N_{\rm{s}}$ data streams, the transmitter and receiver contain $N_{\rm{t}}^{\rm{RF}}$ and $N_{\rm{r}}^{\rm{RF}}$ RF chains, satisfying $N_{\rm{s}} \leq N_{\rm{t}}^{\rm{RF}} \leq N_{\rm{t}}$ and $N_{\rm{s}} \leq N_{\rm{r}}^{\rm{RF}} \leq N_{\rm{r}}$. We assume that the number of RF chains equals the number of data streams for both transmitter and receiver. Thus, the $N_{\rm{s}} \times 1$ symbol vector ${\bf{s}}$ is followed by a $N_{\rm{s}} \times N_{\rm{s}}$ digital precoder ${\bf{F}}_{\rm{D}}$ and $N_{\rm{t}} \times N_{\rm{s}}$ analog precoder ${\bf{F}}_{\rm{A}}$. The received signal is expressed as
\setcounter{equation}{0}
\begin{equation}
\label{received signal}
\begin{aligned}
{\bf{y}} = {\bf{H}} {\bf{F}}_{\rm{A}} {\bf{F}}_{\rm{D}} {\bf{s}} + {\bf{n}},
\end{aligned}
\end{equation}
where ${\bf{y}}$ denotes the $N_{\rm{r}} \times 1$ received vector, ${\bf{H}}$ is the $N_{\rm{r}} \times N_{\rm{t}}$ channel matrix, and ${\bf{n}}$ represents the noise vector ${\bf{n}}=\left[n_1, n_2, \cdots, n_{N_{\rm{r}}}\right]$, whose entities are mutually independent and follow the complex Gaussian distribution $\mathcal{CN}(0,\sigma_{\rm{n}}^2)$.
\par
Due to the high attenuation of millimeter-wave and terahertz bands, LoS path plays a dominant role in the XL-MIMO channel. Therefore, the classical far-field channel can be approximated by \cite{Ayach'14'j}
\begin{equation}
\label{eq: SV-channel}
\begin{aligned}
{\bf{H}} & = \sum\limits_{l = 1}^L  \alpha_l \Lambda_{\rm{r}}(\phi_l^{\rm{r}}) \Lambda_{\rm{t}}(\phi_l^{\rm{t}})  {\bf{a}}_{\rm{r}}(\phi_l^{\rm{r}}) {\bf{a}}_{\rm{t}}^H(\phi_l^{\rm{t}}) \\ 
& \approx \alpha_{\rm{LoS}} \Lambda_{\rm{r}}(\phi_{\rm{LoS}}^{\rm{r}}) \Lambda_{\rm{t}}(\phi_{\rm{LoS}}^{\rm{t}}) {\bf{a}}_{\rm{r}}(\phi_{\rm{LoS}}^{\rm{r}}) {\bf{a}}_{\rm{t}}^H(\phi_{\rm{LoS}}^{\rm{t}}),
\end{aligned}
\end{equation}
where $\alpha_{\rm{LoS}}$ $(\alpha_l)$ is the gain of the LoS $(l^{th})$ path, $\phi_{\rm{LoS}}^{\rm{t}}$ $(\phi_l^{\rm{t}})$ and $\phi_{\rm{LoS}}^{\rm{r}}$ $(\phi_l^{\rm{r}})$ are the azimuth angles of departure and arrival (AoDs/AoAs) for the LoS $(l^{th})$ path, respectively. The functions $\Lambda_{\rm{t}}(\phi_{\rm{LoS}}^{\rm{t}})$ and $\Lambda_{\rm{r}}(\phi_{\rm{LoS}}^{\rm{r}})$ represent the antenna element gain for the transmitter and receiver at the corresponding angle, respectively. The antenna array response vectors ${\bf{a}}_{\rm{t}}$ and ${\bf{a}}_{\rm{r}}$ are introduced to capture the spatial correlation characteristics for MIMO communications. For an $N$-element uniform linear array (ULA), the far-field array response vector ${\bf{a}}(\phi)$ can be written as
\begin{equation}
\label{ULA array response vector}
\begin{aligned}
{\bf{a}}(\phi) = \frac{1}{\sqrt{N}} [1, e^{j \frac{2\pi}{\lambda}d {\rm{sin}}\phi}, \cdots, e^{j (N-1)\frac{2\pi}{\lambda}d{\rm{sin}}\phi}]^T,
\end{aligned}
\end{equation}
where $d$ denotes the element spacing, and $\lambda$ denotes the carrier wavelength. Note that the far-field channel in (\ref{eq: SV-channel}) has a low rank, i.e., it has very limited DoFs. Therefore, with much fewer RF chains than the number of antennas, the classical hybrid precoding architecture can fully utilize the limited spatial DoFs to achieve the near-optimal spectrum efficiency.
\begin{figure}[!t]
	\centering
	\setlength{\abovecaptionskip}{0.cm}
	\includegraphics[width=3in]{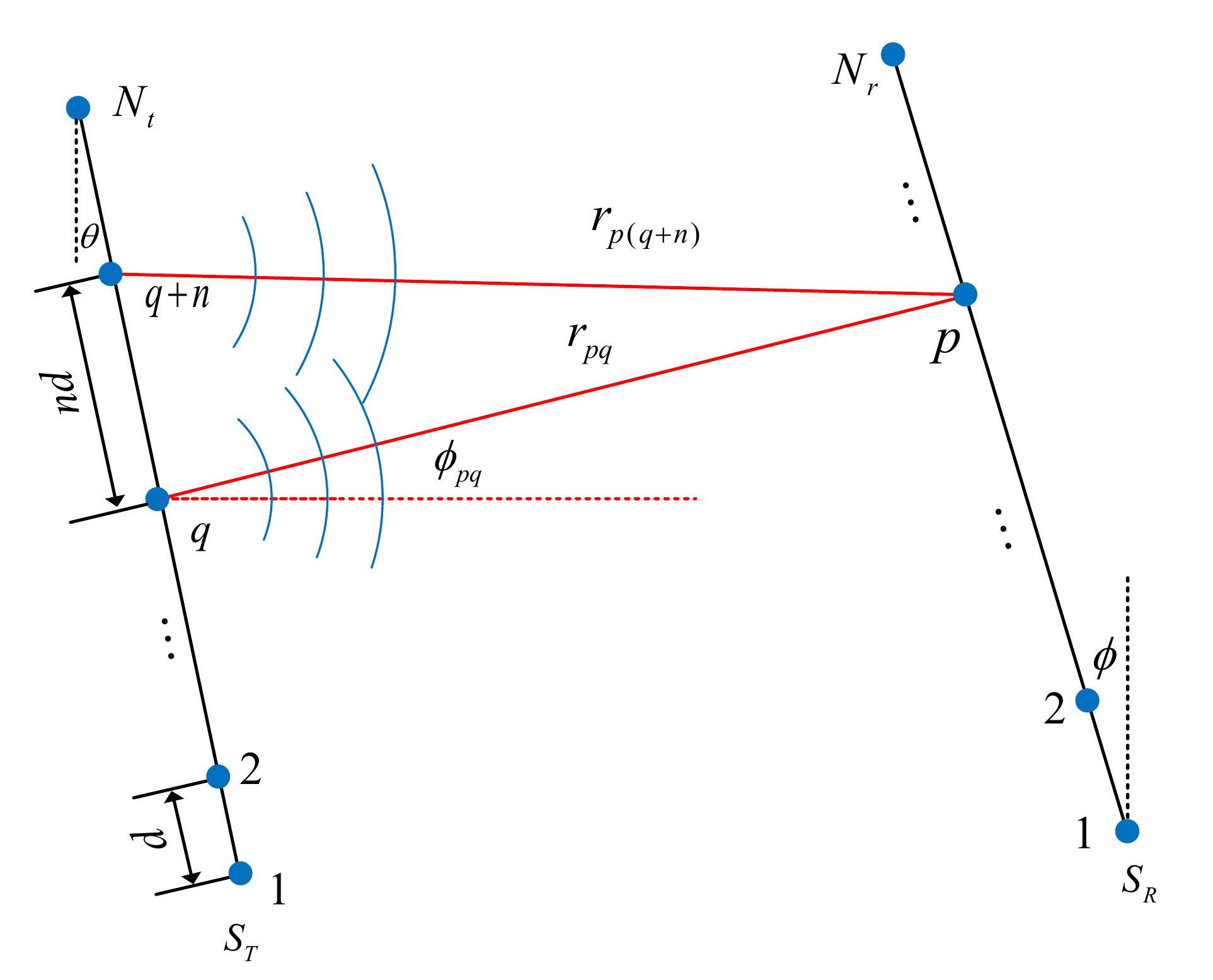}
	\caption{The near-field geometric channel model of two non-parallel arrays.}
	\label{img:near-field}
	\vspace{-0.5cm}
\end{figure}
\par
However, with the extremely large number of antennas in both the transmitter and receiver in XL-MIMO, receivers may lie in the near-field region of the transmitter. Then, the far-field LoS channel based on planar wave assumptions in (\ref{eq: SV-channel}) is no longer accurate to describe the LoS channel in the near-field region, where the channel has to be modeled by spherical wave assumptions. As shown in Fig. \ref{img:near-field}, based on the geometric free space assumption \cite{Sherman'62'j}, the near-field LoS channel can be modeled as 
\begin{equation}
\label{eq: NF-channel}
{\bf{H}} = 
\left[
\begin{matrix}
\alpha_{11} e^{-j \frac{2\pi}{\lambda} r_{11}}& \cdots & \alpha_{1N_{\rm{t}}} e^{-j \frac{2\pi}{\lambda} r_{1N_{\rm{t}}}} \\
\vdots & \ddots & \vdots \\
\alpha_{N_{\rm{r}}1} e^{-j \frac{2\pi}{\lambda} r_{N_{\rm{r}}1}}& \cdots & \alpha_{N_{\rm{r}}N_{\rm{t}}} e^{-j \frac{2\pi}{\lambda} r_{N_{\rm{r}}N_{\rm{t}}}} \\
\end{matrix}
\right],
\end{equation}
where $r_{pq}$ and $\alpha_{pq}$ denote the distance and normalized path gain between $p^{th}$ receiver antenna and $q^{th}$ transmitter antenna, respectively.
\par
It is worth noting that ULA is considered in this paper for simplicity, and it is straightforward to deploy the more realistic planar arrays and the main results remain the same\footnote{We also assume that the number of antennas for both the transmitter and receiver is large. This scenario can happen in backhauls or internet of vehicles in future 6G.}. \begin{proposition} The far-field channel model in (\ref{eq: SV-channel}) is a particular case of the near-field channel in (\ref{eq: NF-channel}) when the array apertures $(N_{\rm{r}}-1)d$ and $(N_{\rm{t}}-1)d$ can be negligible compared with the communication distance $r$.
\end{proposition}
\begin{IEEEproof}
The proof is provided in Appendix A.
\end{IEEEproof}
\par Moreover, it can be obtained that the rank of the far-field LoS channel in (\ref{eq: SV-channel}) is only $1$. As a contrast, the rank of the near-field LoS channel in (\ref{eq: NF-channel}) can be huge, or the channel can even be full rank, which indicates a significantly increased multiplexing gain compared with the far-field LoS channel. To investigate the capacity gain in the near-field region, the available DoFs need to be approximated.

\section{DoFs and Capacity Analysis in Near-Field}\label{sec:theo}
In this section, the theoretical analysis of DoFs and channel capacity in the near-field region is provided. In Section \ref{sec:theo dof_appro}, we analyze the increased available DoFs by introducing eigenproblems in the electromagnetic wave theory. In Section \ref{sec:theo channel_capacity}, to avoid the complicated process of channel decomposition, an estimation of the channel capacity in the near-field region is provided. Finally, we investigate the capacity curve in the near-field region in Section \ref{sec:theo discussion}.
\subsection{DoFs Estimation of the Near-field LoS Channel}\label{sec:theo dof_appro}
\par It is difficult to directly analyze the DoFs, to be precise, the singular values, of the near-field LoS channel. It has been proved that, through the Nyquist space sampling theorem, the half-wavelength spaced antennas can be equivalently modeled as continuous apertures without loss of accuracy. Therefore, we could generalize the classical discrete antenna arrays to continuous apertures. By incorporating the analysis method in electromagnetic field theory, we are able to estimate the DoFs in the near-field region.
\par  
In \cite{miller'00'j}, the DoFs of a pair of \emph{parallel} positioned continuous linear arrays are researched. However, the assumption of parallel arrays is hard to be guaranteed in real communication systems. Therefore, the DoFs of \emph{non-parallel} linear arrays are needed to be analyzed. In this paper, we consider a pair of \emph{non-parallel} positioned continuous linear arrays\footnote{We assume that the arrays are modeled in 3D space, where the linear array is along with the y-direction with the angle of $\theta$ or $\phi$. The width of the array in other directions is small and can be neglected. Therefore, the current on the antenna array is not restricted along with the array.} as in Fig. \ref{img:near-field}. The angle between the array and vertical line is denoted by $\theta$ and $\phi$ for transmitter and receiver, respectively. The centers of the two arrays are positioned on the same level, with separation distance denoted by $r$. We consider a location vector denoted by ${\bm{r}}$. The area within the transmitter and receiver is denoted by $S_T$ and $S_R$, respectively. Since the arrays are modeled as continuous apertures, we can assume the monochromatic source $\psi({\bm{r}}_{\rm{T}})$ on the transmitter Tx that generate waves $\phi({\bm{r}}_{\rm{R}})$ on the receiver array Rx, which derives from Green’s function as
\begin{equation}
\label{Helmholtz}
\begin{aligned}
\phi({\bm{r}}) = \int_{S_T} G({\bm{r}}, {\bm{r}}_{\rm{T}}) \psi({\bm{r}}_{\rm{T}}) d{\bm{r}}_{\rm{T}},
\end{aligned}
\end{equation}
where $G({\bm{r}}, {\bm{r}}_{\rm{T}})$ denotes the corresponding Green’s function. Based on the spherical wave assumptions in the near-field region, the possible waves resulting from a single point source ${\bm{r}}_1$ in free space can be modeled as
\begin{equation}
\label{Green Function}
\begin{aligned}
G({\bm{r}}, {\bm{r}}_1) = \frac{\exp(-jk|{\bm{r}}-{\bm{r}}_1|)}{4\pi|{\bm{r}}-{\bm{r}}_1|},
\end{aligned}
\end{equation}
where the nagative sign inside the exponential represents the outgoing directions. Similar to singular value decomposition (SVD), we introduce the function as 
\begin{equation}
\label{eq: definition of K 1}
\begin{aligned}
K({\bm{r}}_T', {\bm{r}}_T) = \int_{S_R} G^{*}({\bm{r}}_R, {\bm{r}}_T') G({\bm{r}}_R, {\bm{r}}_T)d{\bm{r}}_R.
\end{aligned}
\end{equation}
Therefore, following the formulation in \cite{miller'00'j}, we could derive a series of eigenproblems as 
\begin{equation}
\label{eq: eigenfunction of K 1}
\begin{aligned}
\upsilon \psi({\bm{r}}_T) = \int_{S_T} K({\bm{r}}_T', {\bm{r}}_T) \psi({\bm{r}}_T') d{\bm{r}}_T',
\end{aligned}
\end{equation}
where $\upsilon$ denotes the eigenvalue corresponding to the function $\psi(\cdot)$. 
\begin{proposition} The solution to the eigenproblem in (\ref{eq: eigenfunction of K 1}) could be further simplified as 
\begin{equation}
\label{eigenfunction prolate spheroidal}
\begin{aligned}
\upsilon_n \psi_n(c_y, \xi_T) = \int_{-1}^{1} \frac{{\rm{sin}}[c_y(\xi_T - \xi_{T}')]}{\pi(\xi_T - \xi_{T}')} \psi_n(c_y, \xi_{T}') d\xi_{T}',
\end{aligned}
\end{equation}
where $\psi_n(c_y, \xi_T)$ represents a series of complete orthonormal basis sets, in which $\xi_{T}$ and $\xi_{T}'$ represent the scaled length with given $\xi_{T} = \eta_{T}/((N_t-1)d\cos\theta)$. The dimensionless parameter $c_y$ represents the equivalent bandwidth along the array aperture defined as 
\begin{equation}
\label{eq: definition of c}
\begin{aligned}
c_y = \frac{\pi(N_t-1)(N_r-1)d^2\cos\theta\cos\phi}{2\lambda r}.
\end{aligned}
\end{equation}
\end{proposition}
\begin{IEEEproof}
The proof is provided in Appendix B.
\end{IEEEproof}
\par Based on the equation in (\ref{eigenfunction prolate spheroidal}), it can be proved that the prolate spheroidal wave functions (PSWFs) are the solutions to the eigenproblem \cite{Slepian'1961'j}. The normalized distinct eigenvalues follow the inequation as
\begin{equation}
\label{inequation eigenvalues}
\begin{aligned}
1 \geq \upsilon_0 > \upsilon_1 > \upsilon_2 > \cdots > 0.
\end{aligned}
\end{equation}
\par

\begin{figure}[!t]
	\centering
	\setlength{\abovecaptionskip}{0.cm}
	\includegraphics[width=3in]{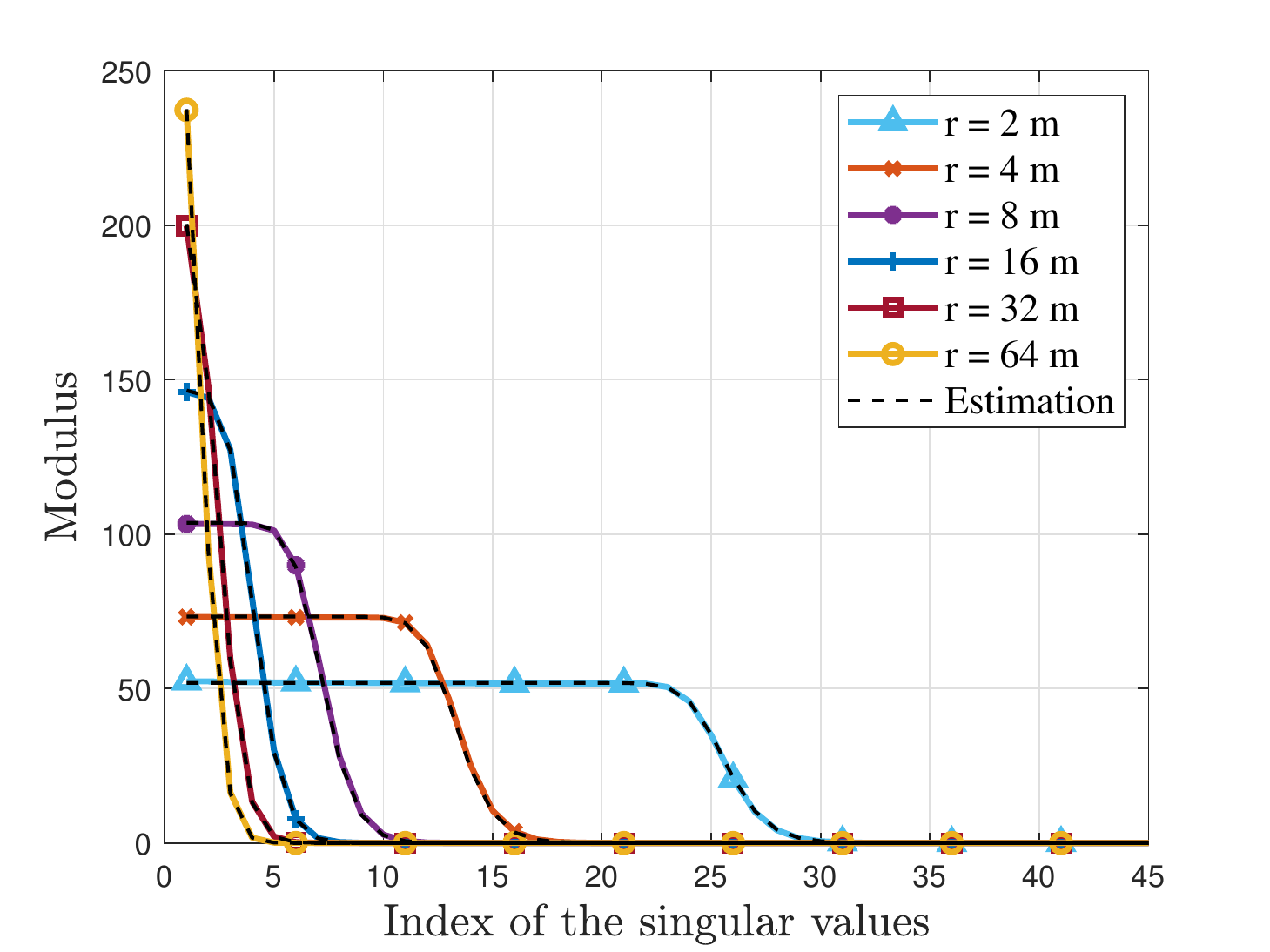}
	\caption{Calculated and estimated singular values for a pair of parallel 256 ULA antennas. Colored solid lines represent the calculated singular values based on SVD on different distances, while black dashed lines are derived with eigenvalues of PSWFs of equal channel power for corresponding distance.}
	\label{img:Singular Values}
	\vspace{-0.5cm}
\end{figure}

In the electromagnetic field theory, the eigenvalues represent the coupling coefficients between orthogonal basis functions, which are equivalent to the square of singular values through SVD in communications. Therefore, the singular values of the near-field LoS channel can be approximated with the eigenvalues of PSWFs. If we consider a pair of parallel arrays, the eigenvalues of PSWFs are consistent with the singular values derived from SVD for a pair of parallel ULA as in Fig. \ref{img:Singular Values}. It indicates that the DoFs can be estimated without complicated SVD when the size of channel matrix ${\bf{H}}$ becomes huge as the number of antennas scales up. It is worth noting that, in the beginning, the eigenvalues fall off slowly. However, there is a critical value after which the eigenvalues of PSWFs (or singular values of the channel) fall off exponentially, which can also be derived with the PSWFs as
\begin{equation}
\label{eq:critical value for rotation}
\begin{aligned}
N_{\rm{DoF}}(r) \approx \frac{2}{\pi}c_y = \frac{(N_{\rm{t}}-1)(N_{\rm{r}}-1)d^2 \cos\theta\cos\phi}{\lambda r}.
\end{aligned}
\end{equation}

As illustrated in (\ref{eq:critical value for rotation}), as the transmitter-receiver distance decreases, the critical value increases, which enables the communication systems to transmit multiple data streams. For special parallel cases, when setting $\theta = 0$ and $\phi = 0$, the estimated DoFs can be simplified as
\begin{equation}
\label{eq:critical value}
\begin{aligned}
N_{\rm{DoF}}(r) \approx \frac{(N_{\rm{t}}-1)(N_{\rm{r}}-1)d^2 }{\lambda r},
\end{aligned}
\end{equation}
which is consistent with the results in \cite{miller'00'j}. Note that $(N_{\rm{t}}-1)d \cos\theta$ represents the effective length of the transmitter array projected along the y-axis. Therefore, a pair of non-parallel arrays is equivalent to the parallel arrays with the effective length projected perpendicularly to the line of array centers for DoFs estimation.
\par
It has been theoretically proved that, unlike classical far-field LoS channel, the spatial DoFs significantly scales up as the distance decreases in the near-field LoS channel. It is expected that channel capacity can be enhanced by utilizing the increasing DoFs for spatial multiplexing in near-field communications.

\subsection{Channel Capacity Analysis}\label{sec:theo channel_capacity}
As mentioned above, the near-field channel differs from the far-field channel in that the DoFs significantly increase within the LoS path. The channel capacity can also benefit from the increased DoFs. If we consider a MIMO system model as in (\ref{received signal}), the channel capacity can be formulated as 
\begin{equation}
\label{eq: capacity}
\begin{aligned}
C &= \mathop {{\rm{max}}}\limits_{{\bf{F}}} ~ {\rm{log}}_2 \left|{\bf{I}}+\frac{1}{\sigma_n^2}{\bf{H}}{\bf{F}}{\bf{F}}^H{\bf{H}}^H \right| \\
&= \sum_{i=1}^{\min(N_t,N_r)} {\rm{log}}_2 \left(1+\frac{p_{i}}{\sigma_n^2}\lambda_i^{2} \right),
\end{aligned}
\end{equation}
where $\lambda_i$ denotes the $i^{th}$ singular value of channel matrix, $p_i$ denotes the power allocated to $i^{th}$ sub-channel according to the classical water-filling method \cite{marzetta'2016'fundamentals}. ${\bf{F}}$ denotes the fully-digital precoding matrix with the power constraints. 
\par
\begin{figure}[!t]
	\centering
	\setlength{\abovecaptionskip}{0.cm}
	\includegraphics[width=3in]{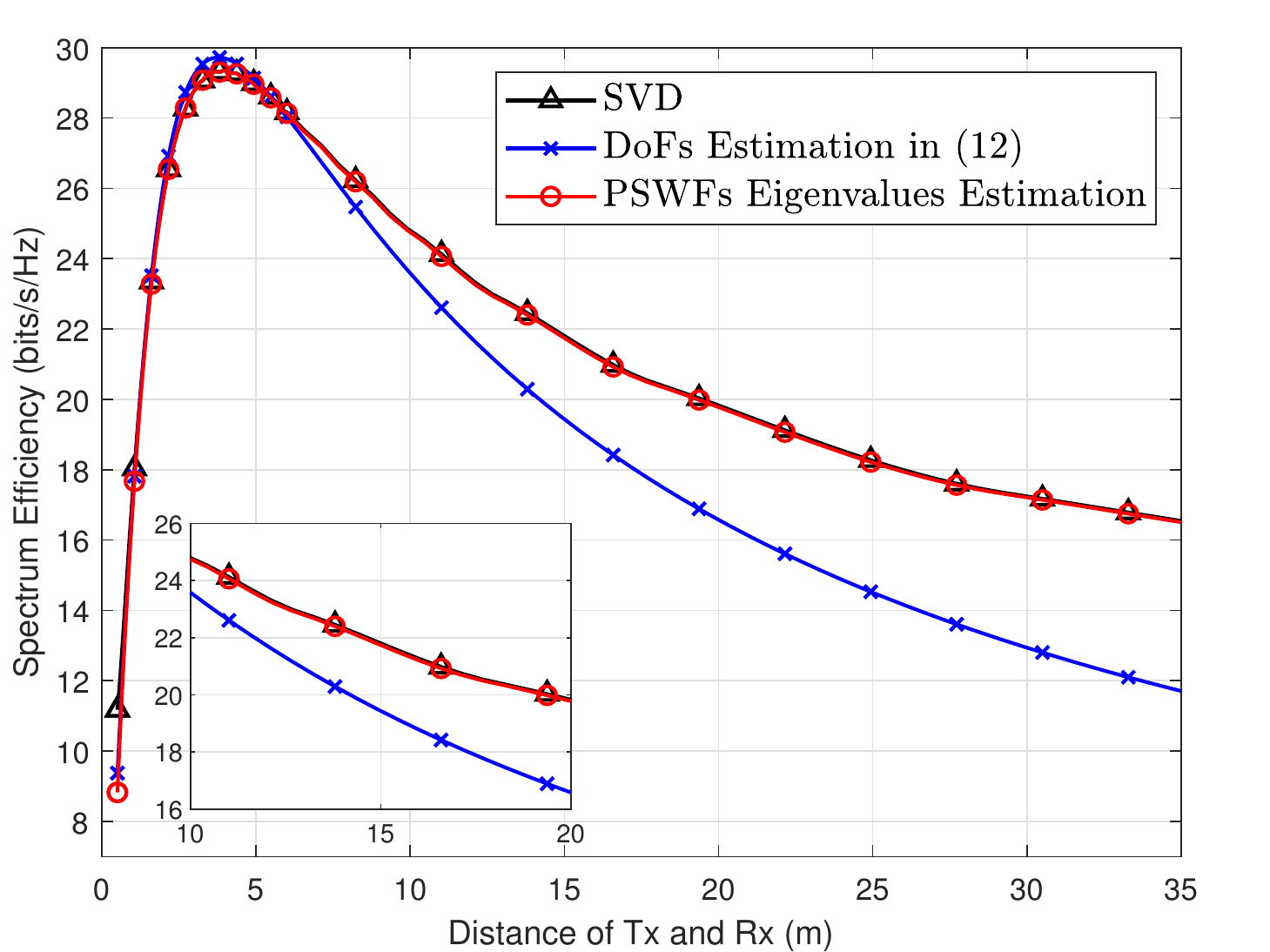}
	\caption{Calculated and estimated channel capacity in the near-field region. Orange line represents the calculated channel capacity, which the purple and red lines denote the estimated channel capacity with DoFs and eigenvalues of PSWFs repectively. The total transmitted SNR is set to be 15 dB.}
	\label{img:capacity curve}
	\vspace{-0.5cm}
\end{figure}

 To find more insights, we consider a case where the array apertures are parallel positioned. Through perfect transmission power control, we could assume the large-scale fading can be ignored, thus the Frobenius norm of the channel matrix share the same value over different distances $r$. According to the estimated singular values of PSWFs in (\ref{eigenfunction prolate spheroidal}), the capacity of the near-field LoS channel can be rewritten as 
\begin{equation}
\label{eq: capacity near-field}
\begin{aligned}
C &\mathop { \approx }\limits^{(a)} \sum_{i=1}^{N_{\rm{DoF}}} {\rm{log}}_2 \left(1+\frac{p_{i}}{\sigma_n^2} \lambda_i^{2} \right) \\
&\mathop { \approx }\limits^{(b)} N_{\rm{DoF}} {\rm{log}}_2 \left(1+\frac{P_{\rm{tot}} P_{\rm{H}}}{\sigma_n^2 N_{\rm{DoF}}^2} \right)\\
& \approx \frac{(N_{\rm{t}}-1)(N_{\rm{r}}-1)d^2}{\lambda r} \\
& \times \log_2 \left(1+\frac{P_{\rm{tot}} P_{\rm{H}}\lambda^2 r^2}{\sigma_n^2 (N_{\rm{t}}-1)^2(N_{\rm{r}}-1)^2d^4} \right),
\end{aligned}
\end{equation}
where $N_{\rm{DoF}}$ denotes the DoFs in the near-field LoS channel as in (\ref{eq:critical value for rotation}), $P_{\rm{H}}$ denotes the power of the channel. Approximation (a) is obtained by assuming that the first $N_{\rm{DoF}}$ sub-channels take up the main power of all the sub-channels. Approximation (b) is obtained by assuming the singular values fall off slowly within the dominant $N_{\rm{DoF}}$ sub-channels and thus equal power is allocated to each sub-channel. According to Fig. \ref{img:Singular Values}, the assumptions above are reasonable for a pair of parallel ULA. Here we have three ways to obtain the channel capacity in the near-field region, by precisely SVD and water-filling method, by assuming an equal power allocation as in (\ref{eq: capacity near-field}) and by water-filling method using eigenvalues of PSWFs. The simulation results are illustrated in Fig. \ref{img:capacity curve}.
\par
It is shown from Fig. \ref{img:capacity curve} that, the capacity significantly increases as the transmission distance decreases. Compared with the classical estimation of DoFs as in \cite{miller'00'j}, the estimation with PSWFs eigenvalues approximates better to the channel capacity for large transmission distance $r$. It reveals a theoretical method to fastly estimate the channel capacity in the near-field region without calculation-consuming SVD process. Although the simulation is carried out under the strict assumptions that the arrays have to be parallel positioned, the method can also be generalized to non-parallel arrays which removes the degeneracy of eigenvalues as illustrated in \cite{miller'19'waves}. The capacity analysis of non-parallel linear arrays in the near-field region is also one direction for future research.
\subsection{Discussions on Channel Capacity}\label{sec:theo discussion}
Based on the approximation results in Fig. \ref{img:capacity curve}, it is worth noting that channel capacity does not monotonically decrease as the distance scales up. We explain it with the approximation result in equation (\ref{eq: capacity near-field}). The denominator within the logarithm contains square of DoFs while the logarithm is also multiplied by DoFs. By derivation, the capacity is maximized when DoFs achieve $N_{\rm{DoF}}^{*}$ which satisfies
\begin{equation}
\label{eq: capacity near-field discuss}
\begin{aligned}
\left( \frac{N_{\rm{DoF}}^{*2} \sigma_n^2}{P_{\rm{tot}}P_{\rm{H}}} +1 \right) {\rm{log}}_2 \left(1+\frac{P_{\rm{tot}}P_{\rm{H}}}{N_{\rm{DoF}}^{*2} \sigma_n^2} \right) = \frac{2}{{\rm{ln}}2}.
\end{aligned}
\end{equation}
Through numerical calculations, the solution to the transcendental equation in (\ref{eq: capacity near-field discuss}) can be obtained by $N_{\rm{DoF}}^{*} \approx \sqrt{0.255\frac{P_{\rm{tot}}P_{\rm{H}}}{\sigma_n^2}}$. The transmitter-receiver distance $r$ can also be estimated with $N_{\rm{DoF}}^{*}$. The special phenomenon originates from the spread of the channel power. For near-field communications, although the DoFs scale up as the distance decreases, the energy of the channel is also averagely allocated to more orthogonal sub-channels, reducing the equivalent signal-noise ratio (SNR) within each sub-channel. Therefore, the capacity for each sub-channel decreases as the DoFs increase, resulting in non-monotonic increasing capacity as distance decreases\footnote{It is worth noting that, here we assume the large-scale fading is not considered in our model. When large-scale fading is incorporated, the capacity will be monotone increasing as the distance decreases.}.

\par
From Fig. \ref{img:Singular Values} and Fig. \ref{img:capacity curve}, as the communication distance decreases, the capacity of near-field LoS channel dramatically (but not monotonically) increases. Therefore, it is expected to utilize the increase DoFs to improve the spectrum efficiency in the near-field region.

\section{Proposed DAP Architecture}\label{sec:architecture} 
\par
With the small DoFs of the far-field LoS channel, the classical far-field hybrid precoding with a reduced number of RF chains could achieve the near-optimal spectrum efficiency. However, with the increased DoFs of the near-field LoS channel, classical hybrid precoding with limited RF chains is unable to approach the optimal spectrum efficiency. According to our best knowledge, an architecture that can flexibly change the number of RF chains to match the increased DoFs in the near-field region is still a blank.
\par
\begin{figure}[!t]
	\centering
	\setlength{\abovecaptionskip}{0.cm}
	\includegraphics[width=3.3in]{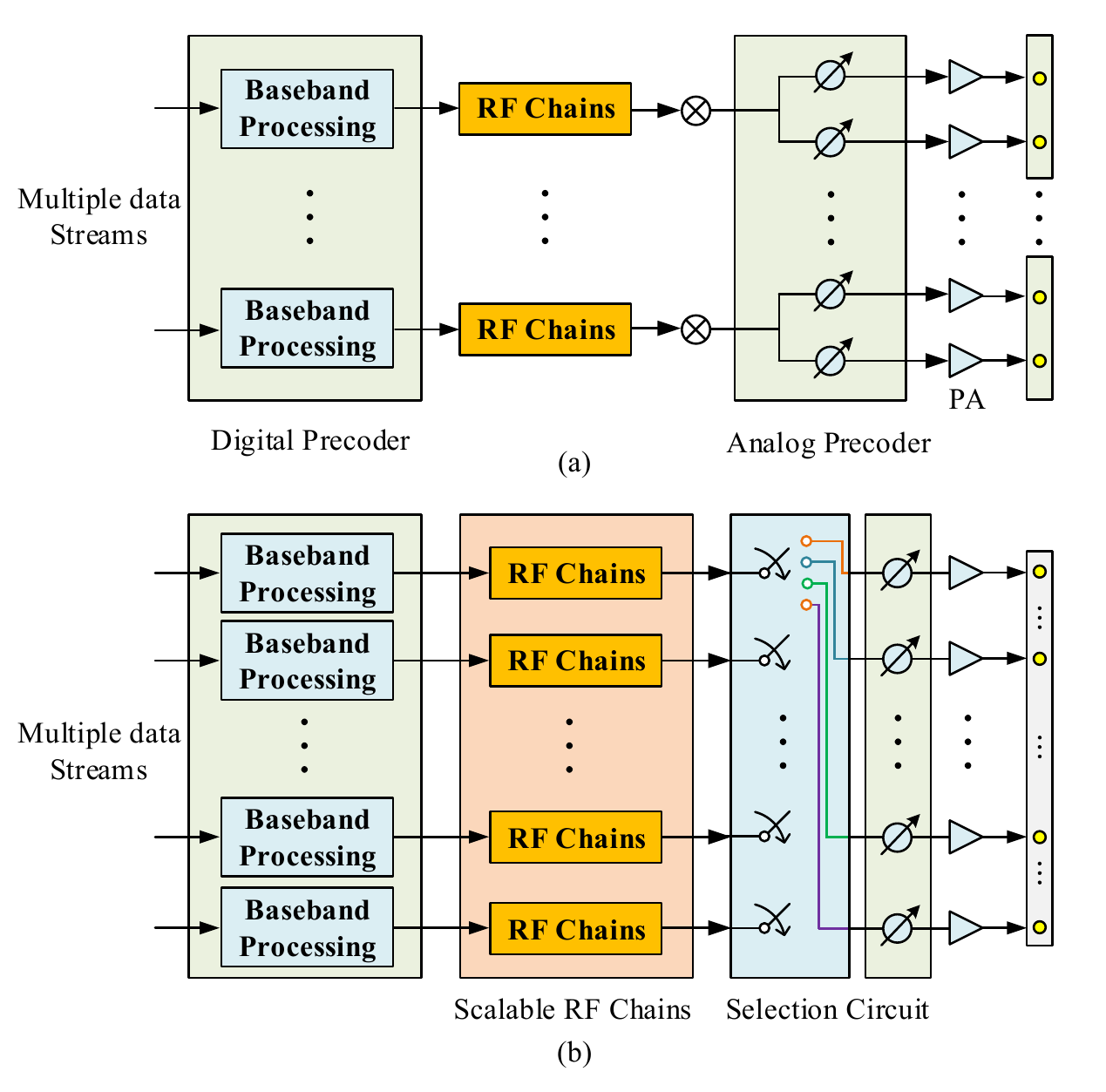}
	\caption{Comparison of the precoding architecture: (a) Classical sub-connected hybrid precoding architecture; (b) Proposed DAP architecture.}
	\label{img:network}
	\vspace{-0.5cm}
\end{figure}

 
To efficiently utilize the DoFs of the near-field LoS channel, fully-digital precoding is able to achieve the optimal spectrum efficiency regardless of in the far-field region or near-field region. Nevertheless, fully-digital precoding requires the number of RF chains equal to the number of transmitter antennas, which leads to the unaffordable power consumption in millimeter-wave and terahertz bands. It is worth noting that the mobile receiver may sometimes be located in the far-field region and be located in the near-field region at other times, resulting in varying distance-related DoFs of the MIMO channel as illustrated in (\ref{eq:critical value for rotation}). Actually, the extra activated RF chains more than the DoFs of the channel can be turned off without significant loss of spectrum efficiency. Motivated by this observation, we propose a DAP architecture that can flexibly control the number of activated RF chains to match the varying DoFs, which can significantly increase the spectrum efficiency in the near-field region. The proposed DAP architecture is shown in Fig. \ref{img:network}.

\par
Specifically, the sub-connected architecture is adopted in the proposed DAP architecture to save unnecessary power consumptions as in \cite{Gao'16'j}. In the DAP architecture, each antenna element only connects to one RF chain. To enable a variable number of RF chains, extra RF chains are equipped in the transmitter and receiver and each RF chain can be configured to active or inactive. In addition, each antenna element can dynamically select one arbitrary RF chain through a selection circuit. The RF chains without any connection can be regarded as inactive, thus enabling the control of the number of RF chains. 
\par The advantage of introducing the selection circuit is two-fold. First, the selection pattern between RF chains and antennas enables the choice of the active RF chain numbers, which can be adaptively adjusted according to the available channel DoFs. Second, the selection pattern also provides a new dimension to design the precoding, which can further enhance the spectrum efficiency.

\par
Based on the proposed DAP structure, the received signal in equation (\ref{received signal}) can be rewritten as
\begin{equation}
\label{received signal rewritten}
\begin{aligned}
{\bf{y}} = {\bf{H}} {\bf{F}}_{\rm{A}} {\bf{F}}_{\rm{S}} {\bf{F}}_{\rm{D}} {\bf{s}} + {\bf{n}},
\end{aligned}
\end{equation}
where the $N_{\rm{t}} \times N_{\rm{s}}$ selection matrix ${\bf{F}}_{\rm{S}}$ represents the adjustable selection matrix between array elements and activated RF chains. Each entity of ${\bf{F}}_{\rm{S}}$ is assigned either $0$ or $1$. Since each antenna element can only select one RF chain, each row of ${\bf{F}}_{\rm{S}}$ contains one element that equals $1$. Besides, since the analog precoder is implemented by analog phase shifters, ${\bf{F}}_{\rm{A}}$ has an extra constant norm constraint. Different from the classical hybrid precoding architecture, the introduction of selection matrix changes the size of analog precoder, which is a $N_{\rm{t}} \times N_{\rm{t}}$ diagonal matrix with constant norm. The feasible set of the analog precoder can be written as
\begin{equation}
\label{feasible set}
\begin{aligned}
{\cal{F}} \triangleq \left\{ {\rm{diag}}(\widetilde{\theta}_{1}, \cdots, \widetilde{\theta}_{N_{\rm{t}}}) ~ \Big| ~\widetilde{\theta}_{i} \in \mathbb{C}, ~\left| \widetilde{\theta}_{i} \right| = 1, ~\forall i \right\}.
\end{aligned}
\end{equation}
\begin{remark}
Unlike the classical hybrid precoding architecture where the number of data streams $N_{\rm{s}}$ is fixed, the proposed DAP architecture could flexibly adjust the number of data streams $N_{\rm{s}}$. Therefore, the optimization of the number of RF chains $N_{\rm{s}}$ should also be taken into account.
\end{remark}
\par When considering the precoding algorithm to maximize the spectrum efficiency, we can assume that an optimal combiner is employed at the receiver side \cite{Ayach'14'j}. Thus, the spectrum efficiency can be written as
\begin{equation}
\label{spectrum efficiency}
\begin{aligned}
R = \log_2\left(\left|{\bf{I}}+\frac{1}{\sigma_{\rm{n}}^2} {\bf{H}}{\bf{F}}_{\rm{A}}{\bf{F}}_{\rm{S}}{\bf{F}}_{\rm{D}}{\bf{F}}_{\rm{D}}^H{\bf{F}}_{\rm{S}}^H{\bf{F}}_{\rm{A}}^H{\bf{H}}^H\right|\right).
\end{aligned}
\end{equation}
Unlike the previous work which assumes an equal power allocation over different sub-channels \cite{Gao'16'j}, we consider a power allocation process according to the classical water-filling method, which will be detailed described in the following section.

\section{Proposed DAP Algorithm}\label{sec:Alg}
In this section, we propose a DAP algorithm for the proposed DAP structure to improve the spectrum efficiency for near-field XL-MIMO communications. In Section \ref{sec:alg overview}, we provide an overview of the DAP algorithm. The design of digital precoder, selection matrix, and analog precoder are introduced from Section \ref{sec:alg ns} to Section \ref{sec:alg analog}.
\subsection{Overview of the DAP Algorithm}\label{sec:alg overview}
Based on the proposed DAP structure, the key problem is to determine the number of data streams $N_{\rm{s}}$. As $N_{\rm{s}}$ increases, the proposed DAP architecture approaches the fully-digital precoding architecture, which is considered to be ideal in terms of spectrum efficiency with high power consumption. According to the analysis in Section \ref{sec:theo}, the DoFs can be well estimated in the near-field region. Similar to the hybrid precoding scheme in the far-field region, a reduced number of RF chains can be deployed to match the DoFs in the near-field region to reduce power consumption as well as achieve the near-optimal spectrum efficiency. With fixed data streams $N_{\rm{s}}$, the optimization can be formulated as
\begin{align}
{{\cal P}:}~&\mathop{{\rm{max}}}\limits_{{\bf{F}}_{\rm{A}},{\bf{F}}_{\rm{S}}, {\bf{F}}_{\rm{D}}} ~ R({\bf{F}}_{\rm{A}},{\bf{F}}_{\rm{S}}, {\bf{F}}_{\rm{D}}) \notag \\
&~~~\,{\rm{s.t.}}~C_1:\| {\bf{F}}_{\rm{A}}{\bf{F}}_{\rm{S}}{\bf{F}}_{\rm{D}} \|_F^2  \le P_{\rm{tot}}\notag\\
&~~~~~~~~\,C_2:{\bf{F}}_{\rm{A}} \in \cal{F}\notag \\
\label{eq: optimization formulation}
&~~~~~~~~\,C_3:({\bf{F}}_{\rm{S}})_{ij} \in \{ 0,1 \},~\forall i,j\\
&~~~~~~~~\,C_4:{\rm{diag}}({\bf{F}}_{\rm{S}}{\bf{F}}_{\rm{S}}^H) = {\bf{1}}_{N_t},\notag
\end{align}
where $\cal{F}$ is the feasible set for analog precoder in (\ref{feasible set}). The precoding algorithm can be summarized into three stages. First, the number of data streams as well as the number of activated RF chains $N_{\rm{s}}$ is determined according to the DoFs for the near-field LoS channel. Second, we determine the selection matrix ${\bf{F}}_{\rm{S}}$ by optimizing the partitioning pattern of the arrays, which aims to choose the best pattern of subarrays that maximizes the spectrum efficiency. Third, we can obtain the corresponding analog precoder ${\bf{F}}_{\rm{A}}$ and digital precoder ${\bf{F}}_{\rm{D}}$ that satisfy the constraints in $C_1$ and $C_2$. The proposed DAP algorithm can be summarized in {\bf Algorithm 1}.

\begin{algorithm}[!t] 
	\caption{Proposed DAP Alogorithm.} 
	\label{alg:1} 
	\begin{algorithmic}[1] 
		\REQUIRE ~ 
		Channel ${\bf{H}}$, $r$, $N_{\rm{t}}$ and $N_{\rm{r}}$.
		\ENSURE ~ 
		Optimized data streams $N_{\rm{s}}^{\rm{opt}}$, digital precoder ${\bf{F}}_{\rm{D}}^{\rm{opt}}$, analog precoder ${\bf{F}}_{\rm{A}}^{\rm{opt}}$ and selection matrix ${\bf{F}}_{\rm{S}}^{\rm{opt}}$
		\STATE Obtain the eigenvalues of PSWFs in (\ref{eigenfunction prolate spheroidal}) with distance $r$;
		\STATE Set the data streams $N_{\rm{s}}^{\rm{opt}}$ by $N_{\rm{DoF}}$ calculated by water-filling method with eigenvalues of PSWFs;
		\STATE Obtain subarray sets $\mathcal{S}_1, \mathcal{S}_2, \cdots, \mathcal{S}_{N_{\rm{s}}^{\rm{opt}}}$ by {\bf Algorithm 2};
		\STATE Calculate ${\bf{F}}_{\rm{S}}^{\rm{opt}}$ according to (\ref{eq: selection matrix from S})
		\STATE Obtain ${\bf{F}}_{\rm{A}}^{\rm{opt}}$ by (\ref{eq: analog opt2});
		\STATE Obtain ${\bf{F}}_{\rm{D}}^{\rm{opt}}$ by (\ref{eq: digital precoder});
		\RETURN $N_{\rm{s}}^{\rm{opt}}$, ${\bf{F}}_{\rm{A}}^{\rm{opt}}$, ${\bf{F}}_{\rm{S}}^{\rm{opt}}$, ${\bf{F}}_{\rm{D}}^{\rm{opt}}$. 
	\end{algorithmic}
\end{algorithm}
\subsection{Estimation of Available Data Streams $N_{\rm{s}}$}\label{sec:alg ns}
The analysis of channel DoFs is provided in Section \ref{sec:theo dof_appro}. In (\ref{eq:critical value for rotation}) and (\ref{eq:critical value}), the DoFs can be estimated well for small distance $r$. However, the eigenvalues of PSWFs offer a better approximation of the singular values of the near-field LoS channel. Therefore, we adopt a water-filling process utilizing the eigenvalues of PSWFs to obtain the DoFs. Precisely, $N_{\rm{DoF}}$ is equal to the number of non-zero power allocated to the sub-channels in the water-filling process. After obtaining $N_{\rm{DoF}}$, similar to the classical hybrid precoding architectures, we set $N_{\rm{s}} = N_{\rm{DoF}}$ to fully utilize the spatial multiplexing gain for capacity improvement in the near-field region at minimum energy consumption. 
\begin{remark}
The choice of $N_{\rm{s}} = N_{\rm{DoF}}$ is not the optimal choice, since the increased number of RF chains could further approach to the fully-digital precoding scheme. Yet it is an essential condition that $N_{\rm{s}} \geq N_{\rm{DoF}}$ to utilize the spatial multiplexing gain. In the following sections, we will compare the performance with the different number of data streams to reveal the rationality of the choice of $N_{\rm{s}}$.
\end{remark}
\subsection{Optimization of Digital Precoder ${\bf{F}}_{\rm{D}}$}\label{sec:alg digital}
With fixed number of data streams $N_{\rm{s}}$, to simplify the formulation in (\ref{eq: optimization formulation}), we define the effective channel ${\bf{H}}_{\rm{e}} \triangleq {\bf{H}} {\bf{F}}_{\rm{A}} {\bf{F}}_{\rm{S}}$. The optimization object can be rewritten as
\begin{equation}
\label{eq: capacity effect formulation}
\begin{aligned}
R =  \log_2\left(\left|{\bf{I}}+\frac{1}{\sigma_{\rm{n}}^2} {\bf{H}}_{\rm{e}} {\bf{F}}_{\rm{D}}{\bf{F}}_{\rm{D}}^H{\bf{H}}_{\rm{e}}^H\right|\right).
\end{aligned}
\end{equation}
By assuming the SVD of effective channel ${\bf{H}}_{\rm{e}} = {\bf{U}}_{\rm{e}}{\bf{\Sigma}}_{\rm{e}}{\bf{V}}_{\rm{e}}^H$, we can obtain the solution of digital precoder
\begin{equation}
\label{eq: digital precoder}
\begin{aligned}
{\bf{F}}_{\rm{D}}^{\rm{opt}} = {\bf{V}}_{\rm{e}}{{\bf{\Gamma}}},
\end{aligned}
\end{equation}
where ${\bf{\Gamma}}$ denotes a diagonal power allocation matrix ${\bf{\Gamma}} = {\rm{diag}}(\sqrt{p_1}, \sqrt{p_2}, \cdots, \sqrt{p_{N_s}})$. Then the spectrum efficiency can be rewritten as 
\begin{equation}
\label{eq: capacity with FD}
\begin{aligned}
R =  \sum_{i=1}^{N_{\rm{s}}} \log_2 \left(1+\frac{\lambda_i^2({\bf{H}} {\bf{F}}_{\rm{A}} {\bf{F}}_{\rm{S}}) p_i}{\sigma_{\rm{n}}^2} \right),
\end{aligned}
\end{equation}
where $\lambda_i^2(\cdot)$ denotes the $i^{th}$ singular value of the argument matrix. And the elements of ${\bf{\Gamma}}$ can be calculated through the water-filling process in which 
\begin{equation}
\label{eq: power allocation}
\begin{aligned}
p_i =  \left(\frac{1}{\mu} - \frac{\sigma_{\rm{n}}^2}{\lambda_i^2({\bf{H}} {\bf{F}}_{\rm{A}} {\bf{F}}_{\rm{S}})} \right)^{+}.
\end{aligned}
\end{equation}
The auxiliary variable $\mu$ is determined by the power constraint that $\sum_i p_i = P_{\rm{tot}}$. Therefore, the digital precoder ${\bf{F}}_{\rm{D}}$ can be easily obtained when analog precoder ${\bf{F}}_{\rm{A}}$ and selection matrix ${\bf{F}}_{\rm{S}}$ are fixed.

\subsection{Optimization of Selection Matrix ${\bf{F}}_{\rm{S}}$}\label{sec:alg selection}
The selection matrix ${\bf{F}}_{\rm{S}}$ represents the linking pattern of the RF chains and array elements. Here we denote the set of array indices that connect to the $i^{th}$ RF chain as
\begin{equation}
\label{eq: array subset}
\begin{aligned}
\mathcal{S}_i = \left\{n_{i1}, n_{i2}, \cdots, n_{|\mathcal{S}_i|} \right\},
\end{aligned}
\end{equation}
where $|\mathcal{S}_i|$ denotes the number of entities in $\mathcal{S}_i$. The selection matrix can be directly derived from $\mathcal{S}_i$ as 
\begin{equation}
[{\bf{F}}_{\rm{S}}]_{j,i}=\left\{
\label{eq: selection matrix from S}
\begin{aligned}
~~1,~~~ j \in \mathcal{S}_i,\\
~~0,~~~ j \notin \mathcal{S}_i.
\end{aligned}
\right.
\end{equation}
To get the optimal partitioning of subarrays, according to Jensen's inequality, the formulation of spectrum efficiency can be further rewritten as
\begin{equation}
\label{eq: approxi jensen}
\begin{aligned}
R & =  \sum_{i=1}^{N_{\rm{s}}} \log_2 \left(1+\frac{\lambda_i^2({\bf{H}} {\bf{F}}_{\rm{A}} {\bf{F}}_{\rm{S}}) p_i}{\sigma_{\rm{n}}^2} \right) \\
&\le {N_{\rm{s}}} \log_2 \left(1 + \frac{1}{{N_{\rm{s}}}} \sum_{i=1}^{N_{\rm{s}}}  \frac{\lambda_i^2({\bf{H}} {\bf{F}}_{\rm{A}} {\bf{F}}_{\rm{S}}) p_i}{\sigma_{\rm{n}}^2} \right).
\end{aligned}
\end{equation}
\par 
As illustrated in Section \ref{sec:theo dof_appro}, the singular values of the near-field LoS channel are close to each other if $N_{\rm{s}}$ satisfies $N_{\rm{s}} \le N_{\rm{DoF}}$, which is consistent with our settings. Thus the approximation is tight for near-field XL-MIMO communication systems. Then, the optimization of selection matrix ${\bf{F}}_{\rm{S}}$ and analog precoder ${\bf{F}}_{\rm{A}}$ can be simplified as
\begin{equation}
\label{eq: opt problem FS}
\begin{aligned}
({\bf{F}}_{\rm{A}}^{\rm{opt}},{\bf{F}}_{\rm{S}}^{\rm{opt}}) = &\mathop{{\rm{argmax}}} \limits_{{\bf{F}}_{\rm{A}},{\bf{F}}_{\rm{S}}} \sum_{i=1}^{N_{\rm{s}}} \lambda_i^2({\bf{H}}{\bf{F}}_{\rm{A}} {\bf{F}}_{\rm{S}}).
\end{aligned}
\end{equation}
\par
Utilizing the feature of ${\bf{F}}_{\rm{S}}$ that all elements in each row of ${\bf{F}}_{\rm{S}}$ are equal $0$ except one element, ${\bf{F}}_{\rm{S}}$ can be reformulated as 
\begin{equation}
\label{eq: FS diag}
\begin{aligned}
{\bf{F}}_{\rm{S}} = {\bf{P}}_{\rm{S}} \widetilde{{\bf{F}}}_{\rm{S}},
\end{aligned}
\end{equation}
where ${\bf{P}}_{\rm{S}}$ is a permutation matrix and $\widetilde{{\bf{F}}}_{\rm{S}}$ follows a block diagonal pattern of 
\begin{equation}
\widetilde{{\bf{F}}}_{\rm{S}} = \left[
\begin{matrix}
\label{eq: FS special pattern}
 {\bf{1}}_{\mathcal{S}_1}      & 0      & \cdots & 0      \\
 0      & {\bf{1}}_{\mathcal{S}_2}      & \cdots & 0      \\
 \vdots & \vdots & \ddots & \vdots \\
 0      & 0      & \cdots & {\bf{1}}_{\mathcal{S}_{N_{\rm{s}}}}      \\
\end{matrix}
\right],
\end{equation}
where ${\bf{1}}_{\mathcal{S}_i}$ denotes a vector with $|\mathcal{S}_i|$ elements that are equal to 1. Due to the property of permutation matrix, we can switch the order of the diagonal analog precoder ${\bf{F}}_{\rm{A}}$ and permutation matrix ${\bf{P}}_{\rm{S}}$ as
\begin{equation}
\label{eq: FS diag2}
\begin{aligned}
{\bf{H}}{\bf{F}}_{\rm{A}} {\bf{P}}_{\rm{S}}\widetilde{{\bf{F}}}_{\rm{S}} = {\bf{H}}{\bf{P}}_{\rm{S}} \widetilde{{\bf{F}}}_{\rm{A}}\widetilde{{\bf{F}}}_{\rm{S}},
\end{aligned}
\end{equation}
where $\widetilde{{\bf{F}}}_{\rm{A}}$ denotes the diagonal analog precoder with a permutation operation. A brief proof is provided in Appendix C.
\par
Due to the block diagonal feature of $\widetilde{{\bf{F}}}_{\rm{S}}$, $\widetilde{{\bf{H}}} = {\bf{H}} {\bf{P}}_{\rm{S}}$ can be classified by columns as submatrices $\widetilde{{\bf{H}}} = [{\bf{H}}_{\mathcal{S}_1}, \cdots, {\bf{H}}_{\mathcal{S}_{\rm{Ns}}}]$. Therefore, the optimization object in (\ref{eq: opt problem FS}) can be rewritten as
\begin{equation}
\label{eq: opt problem FS2}
\begin{aligned}
&~~~\mathop{{\rm{max}}} \limits_{{\bf{F}}_{\rm{A}},{\bf{F}}_{\rm{S}}} \sum_{i=1}^{N_{\rm{s}}} \lambda_i^2({\bf{H}}{\bf{F}}_{\rm{A}} {\bf{F}}_{\rm{S}}) \\
&= \mathop{{\rm{max}}} \limits_{{\bf{F}}_{\rm{A}},{\bf{F}}_{\rm{S}}} \sum_{i=1}^{N_{\rm{s}}} \lambda_i^2({\bf{H}} {\bf{P}}_{\rm{S}}\widetilde{{\bf{F}}}_{\rm{A}} \widetilde{{\bf{F}}}_{\rm{S}}) \\
&= \mathop{{\rm{max}}} \limits_{{\bf{F}}_{\rm{A}},{\bf{F}}_{\rm{S}}} \left\|{\bf{H}} {\bf{P}}_{\rm{S}}\widetilde{{\bf{F}}}_{\rm{A}} \widetilde{{\bf{F}}}_{\rm{S}}\right\|_F^2 \\
&= \mathop{{\rm{max}}} \limits_{{\bf{F}}_{\rm{A}},{\bf{F}}_{\rm{S}}} \left\|\left[{\bf{H}}_{\mathcal{S}_1}{\bf{f}}_1, \cdots, {\bf{H}}_{\mathcal{S}_{N_{\rm{s}}}}{\bf{f}}_{N_{\rm{s}}} \right]\right\|_F^2 \\
&\mathop { \approx }\limits^{(a)} \sum_{i=1}^{N_{\rm{s}}} \lambda_1({\bf{H}}_{\mathcal{S}_i}^H{\bf{H}}_{\mathcal{S}_i}),
\end{aligned}
\end{equation}
where ${\bf{f}}_i$ denotes non-zero elements of the $i^{th}$ colomn of $\widetilde{{\bf{F}}}_{\rm{A}} \widetilde{{\bf{F}}}_{\rm{S}}$. The approximation (a) is derived by omitting the constant modulus constraints of ${\bf{F}}_{\rm{A}}$. Therefore, the optimization object can be formulated as the summation of the largest singular value of submatrices. However, if we perform exhaustive searching to find all the sets $\mathcal{S}_i$ that maximize the spectrum efficiency, the complexity will be unaffordable. Therefore, we propose a low-complexity algorithm to optimize the selection matrix ${\bf{F}}_{\rm{S}}$ by utilizing the normalized Minkowski $\ell_1$-norm. 
\par
It has been proved that Minkowski $\ell_1$-norm gives a good approximation of the largest singular value \cite{wolkowicz'80'} as
\begin{equation}
\begin{aligned}
\label{eq: Minkowski}
\lambda_1({\bf{R}}_{\mathcal{S}}) \approx \hat{\lambda}_1 ({\bf{R}}, {\mathcal{S}}) = \frac{1}{|\mathcal{S}|}\sum_{i\in \mathcal{S}} \sum_{j\in \mathcal{S}} |[{\bf{R}}_{i,j}]|.
\end{aligned}
\end{equation}
Therefore, we could perform a low-complex greedy searching process to maximize the average Minkowski $\ell_1$-norm for each partitioning set $\mathcal{S}_i$. The proposed near-field subarray partitioning algorithm is summarized in {\bf Algorithm 2}.
\begin{algorithm}[!t] 
	\caption{Near-Field Subarray Partitioning Alogorithm.} 
	\label{alg:2} 
	\begin{algorithmic}[1] 
		\REQUIRE ~ 
		Channel ${\bf{H}}$, $N_{\rm{s}}$, $N_{\rm{bound}}$ and $N_{\rm{t}}$.
		\ENSURE ~ 
		$\mathcal{S}_1, \mathcal{S}_2, \cdots, \mathcal{S}_{N_{\rm{s}}}$
		\STATE ${\bf{R}} = {\bf{H}}^H{\bf{H}}$, $\mathcal{S}_{\rm{sel}} = \varnothing$, $n_{\rm{group}}= \lfloor \frac{N_{\rm{t}}}{N_{\rm{s}}} \rfloor$ 
		\STATE Initialize $\mathcal{S}_i = \{i \cdot n_{\rm{group}}\}$, $\mathcal{S}_{\rm{sel}} \gets \mathcal{S}_{\rm{sel}} \cup \{i \cdot n_{\rm{group}}\}$, for $i=1,2,\cdots,N_{\rm{s}}$
		\FOR{$k = 1:N_{\rm{t}}- N_{\rm{s}}$}
		\STATE $\{ i_k, j_k\} = \mathop{{\rm{argmax}}} \limits_{i \in \mathcal{S}_{\rm{sel}}, j \notin \mathcal{S}_{\rm{sel}}} |[{\bf{R}}]_{i,j}|$
		\STATE $\hat r = \mathop{{\rm{argmax}}} \limits_{r \in \{1,\cdots,N_{\rm{s}} \}} \hat{\lambda}_1({\bf{R}}, \mathcal{S}_{r} \cup \{j_k\}) - \hat{\lambda}_1({\bf{R}},\mathcal{S}_{r})$
		\STATE $\mathcal{S}_{\rm{sel}} \gets \mathcal{S}_{\rm{sel}} \cup j_k$, $\mathcal{S}_{\hat r} \gets \mathcal{S}_{\hat r} \cup j_k$
		\IF{$ |\mathcal{S}_{\hat r}| \geq N_{\rm{bound}}$}
		\STATE $\hat m = \mathop{{\rm{argmin}}} \limits_{m \in \mathcal{S}_{\hat r} } \sum_{n \in \mathcal{S}_{\hat r}} |{\bf{R}}_{m,n}|$
		\STATE $\hat{r}' = \mathop{{\rm{argmax}}} \limits_{r' \neq {\hat{r}}} \hat{\lambda}_1({\bf{R}},\mathcal{S}_{r'} \cup \hat{m}) - \hat{\lambda}_1({\bf{R}},\mathcal{S}_{r'})$
		\STATE $\mathcal{S}_{\hat r} \gets \mathcal{S}_{\hat r} \backslash \hat{m}$, $\mathcal{S}_{\hat{r}'} \gets \mathcal{S}_{\hat{r}'} \cup \hat{m}$
		\ENDIF
		\ENDFOR
		\FOR{$l = 1:N_{\rm{s}}$}
		\STATE $\hat m = \mathop{{\rm{argmin}}} \limits_{m \in \mathcal{S}_{l} } \sum_{n \in \mathcal{S}_{l}} |{\bf{R}}_{m,n}|$
		\STATE $\hat{r}' = \mathop{{\rm{argmax}}} \limits_{r'} \hat{\lambda}_1({\bf{R}},\mathcal{S}_{r'} \cup \hat{m}) - \hat{\lambda}_1({\bf{R}},\mathcal{S}_{r'})$
		\STATE $\mathcal{S}_{\hat r} \gets \mathcal{S}_{\hat r} \backslash \hat{m}$, $\mathcal{S}_{\hat{r}'} \gets \mathcal{S}_{\hat{r}'} \cup \hat{m}$
		\ENDFOR
		\RETURN $\mathcal{S}_1, \mathcal{S}_2, \cdots, \mathcal{S}_{N_{\rm{s}}}$ 
	\end{algorithmic}
\end{algorithm}
\par
Different from other dynamic subarray design algorithms in \cite{Sun'18'c,park'17'j,Xu'19'c}, we aim to obtain subarrays with similar scales. As illustrated in Section \ref{sec:theo dof_appro}, the singular values of the near-field LoS channel are close to each other. Thus the scale of subarrays should be similar to generate equal beamforming gain. First, we initialize all the sets ${\mathcal{S}}$ with uniformly distributed antennas. Then, we add the antennas to sets one by one. We select the $j^{th}$ antenna with the largest absolute value of $|{\bf{R}}_{ij}|$ in which $j^{th}$ antenna is not added yet. If the cardinal of sets exceeds the maximum scale $N_{\rm{bound}}$, we remove the $m^{th}$ antenna with the least contribution and add it to the set with largest improvement. Finally, we check all the sets one by one to remove the least contributor, which aims to eliminate the influence of the manual initialization. 
\begin{remark}
It is worth noting that, as the number of data streams $N_{\rm{s}}$ increases, the heuristic dynamic subarray searching algorithm proposed in \cite{Sun'18'c,park'17'j} may result in a situation where some subarrays are totally empty. The degradation of subarrays causes inefficient usage of spatial DoFs in the near-field region, making the spectrum efficiency far less than optimal. Another greedy searching algorithm proposed in \cite{Xu'19'c} aims to maximize the largest singular value of each submatrix one by one. In this way, the submatrices can hardly remain balanced to maximize the sum rate of all sub-channels. Therefore, the proposed partitioning algorithm is more suitable for near-field communication scenarios.
\end{remark}
\par 
Since in each iteration, one antenna is added into one set, the convergence of the proposed algorithm is guaranteed. In the worst case where one set quickly achieves the maximum size, the complexity of the algorithm is estimated as $\mathcal{O}(N_{\rm{t}}^2 + N_{\rm{t}}N_{\rm{bound}})$ as for comparing operations and $\mathcal{O}(N_{\rm{s}}N_{\rm{t}}N_{\rm{bound}})$ as for add operations.

\subsection{Optimization of Analog Precoder ${\bf{F}}_{\rm{A}}$}\label{sec:alg analog}
As illustrated in (\ref{eq: opt problem FS2}), the optimal analog precoder corresponding to the $i^{th}$ subarray can be formulated as 
\begin{equation}
\label{eq: analog opt}
\begin{aligned}
\angle({\bf{f}}_{{\rm{A}},\mathcal{S}_i}^{\rm{opt}}) = \angle({\bf{v}}_{\mathcal{S}_i}),
\end{aligned}
\end{equation}
where ${\bf{v}}_{\mathcal{S}_i}$ is the singular vector corresponding to the largest singular value of ${\bf{H}}_{\mathcal{S}_i}$. Therefore, each array ${\bf{f}}_{{\rm{A}},\mathcal{S}_i}^{\rm{opt}}$ can be determined satisfying the constraints on the norm. Finally, the formulation of the analog precoder can be written as
\begin{equation}
\label{eq: analog opt2}
\begin{aligned}
{\bf{F}}_{\rm{A}}^{\rm{opt}} = {\bf{P}}_{\rm{S}} {\rm{diag}} \left(\left[({\bf{f}}_{{\rm{A}},\mathcal{S}_1}^{\rm{opt}})^T, \cdots, ({\bf{f}}_{{\rm{A}},\mathcal{S}_{\rm{Ns}}}^{\rm{opt}})^T \right] \right) {\bf{P}}_{\rm{S}}^{T}.
\end{aligned}
\end{equation}

It is worth noting that, due to the constant norm constraints of ${\bf{F}}_{\rm{A}}$, the close-form solution solution in (\ref{eq: opt problem FS2}) is an approximation of the optimal solution without constraints. However, the approximation often induces little performance degradation.


\addtolength{\topmargin}{0.01in}
\section{Simulation Results}\label{sec:simulation}
In this section, to verify the effectiveness of the proposed DAP architecture, we evaluate the performance of spectrum efficiency as well as the energy efficiency compared with classical sub-connected hybrid precoding and classical fully-connected precoding algorithms. Moreover, we investigate the performance for different choices of data streams $N_{\rm{s}}$ in the near-field XL-MIMO communication systems. 
\par 
In the simulations, we consider a single-user XL-MIMO communication scenario. Both the transmitter and receiver are equipped with ULA with $256$ elements spacing of half wavelength. The carrier wave is set to $100$ GHz, with wave length of $\lambda = 3$ mm. It can be calculated that the Rayleigh distance is around $100$ m under the settings. The transmission distance varies within the Rayleigh distance in our simulations. For the wireless channel, as mentioned before, we adopt the deterministic LoS channel in (\ref{eq: NF-channel}) and assume that the large-scale channel fading can be ignored through perfect power control.

\subsection{Evaluating the Performance of Spectrum Efficiency}\label{sec:sim spe_eff}
\par
\begin{figure}[!t]
	\centering
	\setlength{\abovecaptionskip}{0.cm}
	\includegraphics[width=3in]{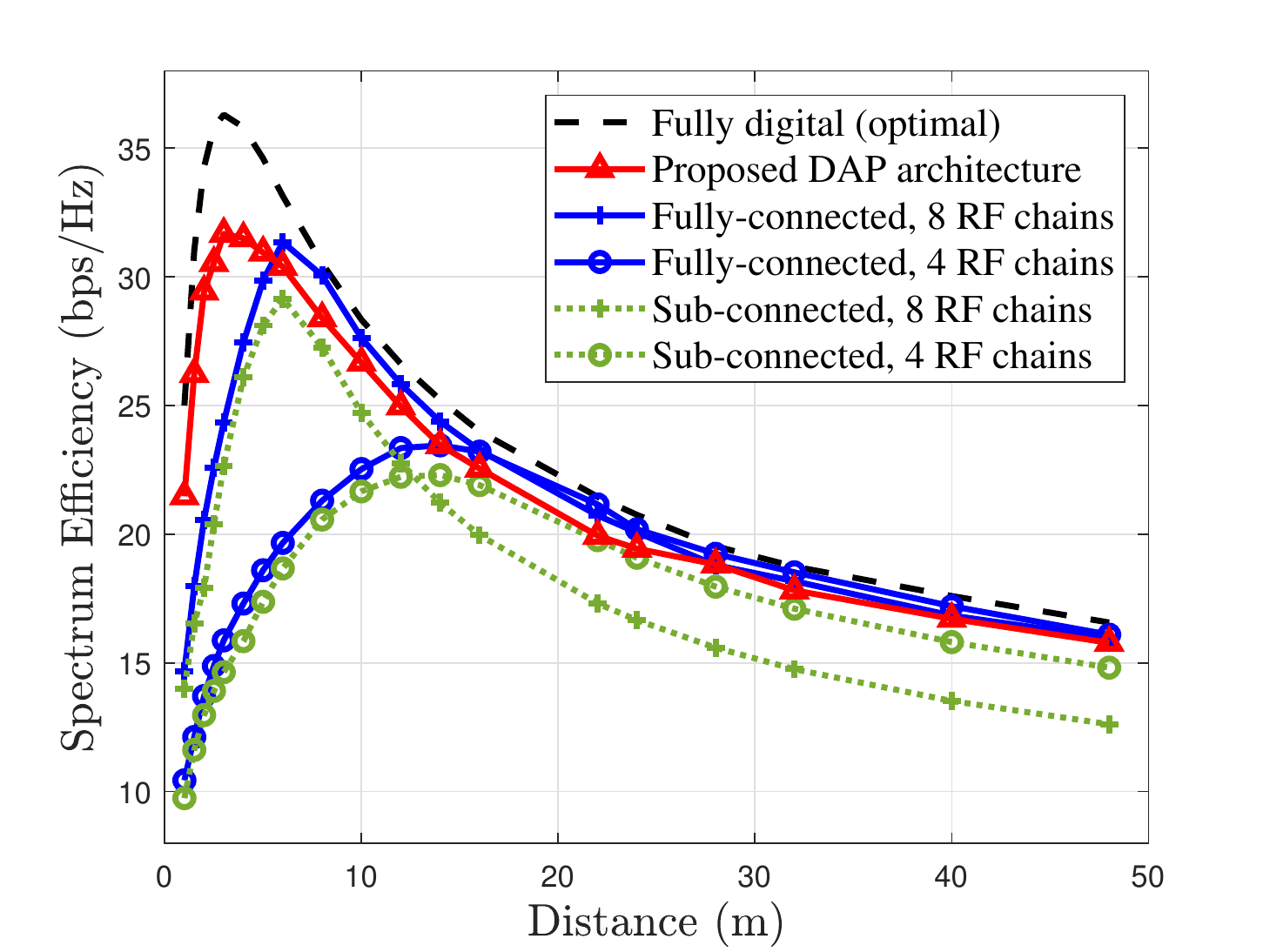}
	\caption{The comparison of spectrum efficiency over different distances.}
	\label{img:SE}
\end{figure}

 The spectrum efficiency for the proposed DAP architecture over different distances is shown in Fig. \ref{img:SE}. The SNR $= P_{\rm{tot}}/\sigma_n^2$ is set to $30$ dB. The baselines include fully-digital precoding, fully-connected hybrid precoding with 8, 4 RF chains using an alternating minimization algorithm\cite{Yu'16'j}, and sub-connected hybrid precoding with 8, 4 RF chains using the SIC-based algorithm\cite{Gao'16'j}. When the transmitter-receiver distance is small, the proposed scheme can achieve about $40\%$ increase in the achievable spectrum efficiency compared to classical hybrid precoding schemes with 8 RF chains. This is because the proposed architecture can benefit from the extra DoFs in the near-field region. When the transmitter-receiver distance grows, the fully-connected hybrid precoding outperforms our proposed architecture. The reason lies in that there still exists a gap between the fully-connected architecture and sub-connected architecture with the same number of RF chains, as illustrated in Section \ref{sec:sim ns}. When the transmitter-receiver distance grows larger, the DoFs reduce and our proposed architecture could no longer obtain the multiplexing gain from the extra RF chains.
\begin{figure}[!t]
	\centering
	\setlength{\abovecaptionskip}{0.cm}
	\includegraphics[width=3in]{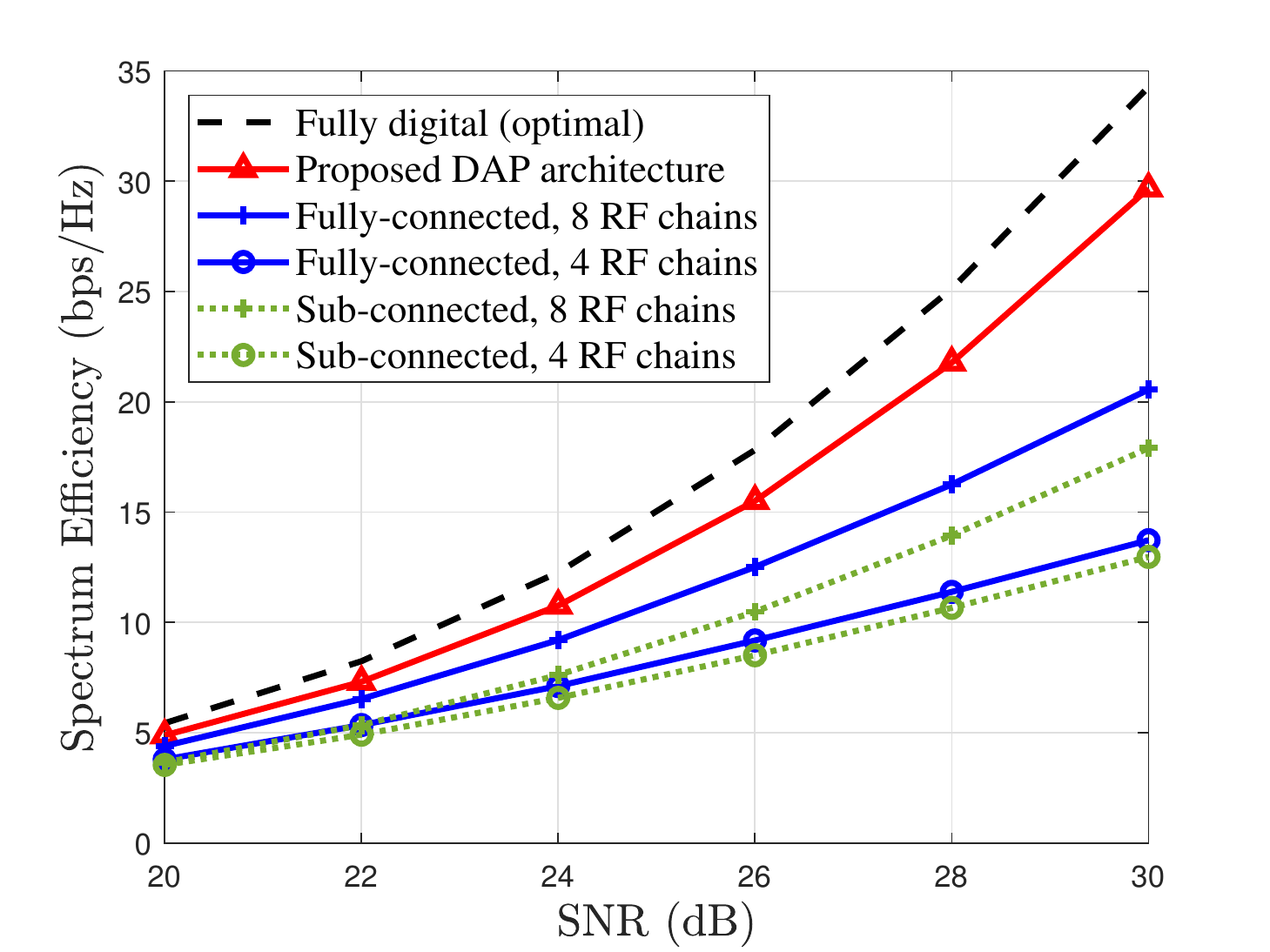}
	\caption{The comparison of spectrum efficiency over different SNRs.}
	\label{img:SE_SNR}
\end{figure}
\par
Another simulation is carried out over different SNRs in Fig. \ref{img:SE_SNR}. The SNR ranges from $20$ dB to $30$ dB. From Fig. \ref{img:SE_SNR}, the proposed DAP architecture outperforms other precoding architectures with fixed RF chains.

\subsection{Evaluating the Performance of Energy Efficiency}\label{sec:sim ene_eff}
\begin{figure}[!t]
	\centering
	\setlength{\abovecaptionskip}{0.cm}
	\includegraphics[width=3in]{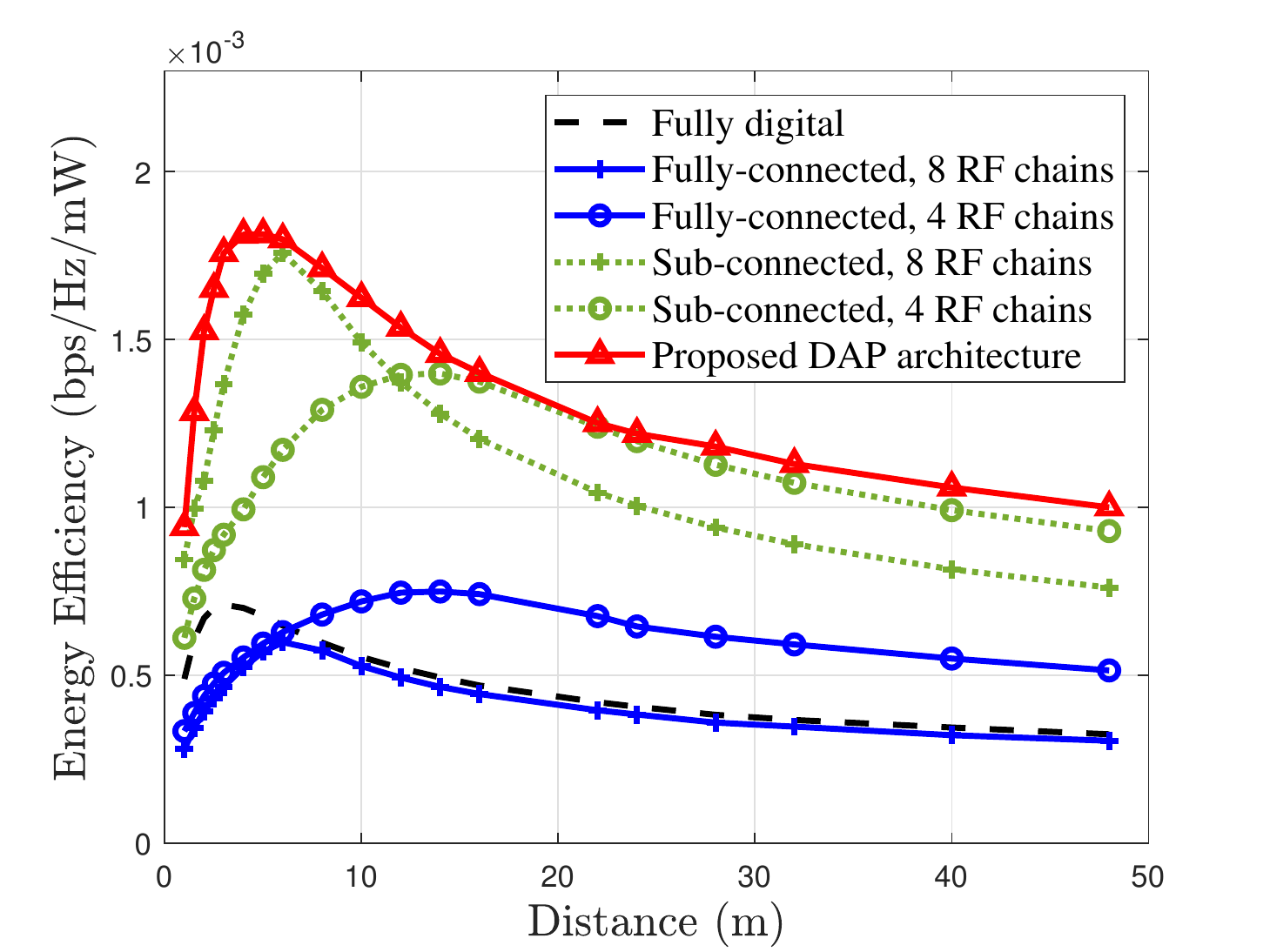}
	\caption{The comparison of energy efficiency over different distances.}
	\label{img:EE}
\end{figure}

With the increasing $N_{\rm{s}}$, the proposed DAP is approaching the classical fully-digital precoding scheme. The advantage of the DAP architecture lies in that it significantly increases the spectrum efficiency utilizing the enlarging DoFs in the near-field region. Also, the scalable RF chains can match the DoFs to save unnecessary energy consumption. Therefore, another critical performance factor, energy efficiency, needs to be considered in the simulations. For simplicity, we claim a downlink power consumption model at the base station, and the energy efficiency can be formulated as
\begin{equation}
\label{power model}
\begin{aligned}
\eta = \frac{R}{P_{\rm{T}} + N_{\rm{s}}P_{\rm{RF}} + N_{\rm{PS}}P_{\rm{PS}} +  N_{\rm{SW}}P_{\rm{SW}} + N_{\rm{t}}P_{\rm{PA}} },
\end{aligned}
\end{equation}
where $P_{\rm{T}}$ is the constant circuit power, $P_{\rm{RF}}$, $P_{\rm{PS}}$ and $P_{\rm{SW}}$ denote the power consumption for each RF chain, phase shifter and switch respectively, $N_{\rm{PS}}$ and $N_{\rm{SW}}$ denote the number of phase shifters and switches respectively. $P_{\rm{T}} = 2500$ mW, $P_{\rm{RF}} = 160$ mW, $P_{\rm{PS}} = 10$ mW, $P_{\rm{SW}} = 10$ mW and $P_{\rm{PA}} = 30$ mW are assumed. The baselines also include the fully-digital precoding, fully-connected and sub-connected hybrid precoding schemes. In fully-connected hybrid precoding architectures, the number of phase shifters is equal to $N_{\rm{PS}}^{\rm{FC}} = N_{\rm{t}}N_{\rm{RF}}$. While in sub-connected hybrid precoding architectures, the number of phase shifters equals to the number of antennas $N_{\rm{PS}}^{\rm{SC}} = N_{\rm{t}}$. Classical fully-connected and sub-connected hybrid architectures are implemented without switches, while in the proposed DAP architectures $N_{\rm{SW}} = N_{\rm{s}}$.  The simulation results of energy efficiency are shown in Fig. \ref{img:EE}. We can observe from Fig. \ref{img:EE} that, the proposed DAP architecture also outperforms all other precoding architectures, which reveals the superiority of the proposed DAP architecture in terms of the energy consumption for transmitting the same amount of data.

\subsection{Evaluating the Effect of Different $N_{\rm{s}}$}\label{sec:sim ns}
\par
\begin{figure}
	\centering
	\subfigure[Spectrum efficiency]{
		\label{img:ns_se}
		\includegraphics[width=3in]{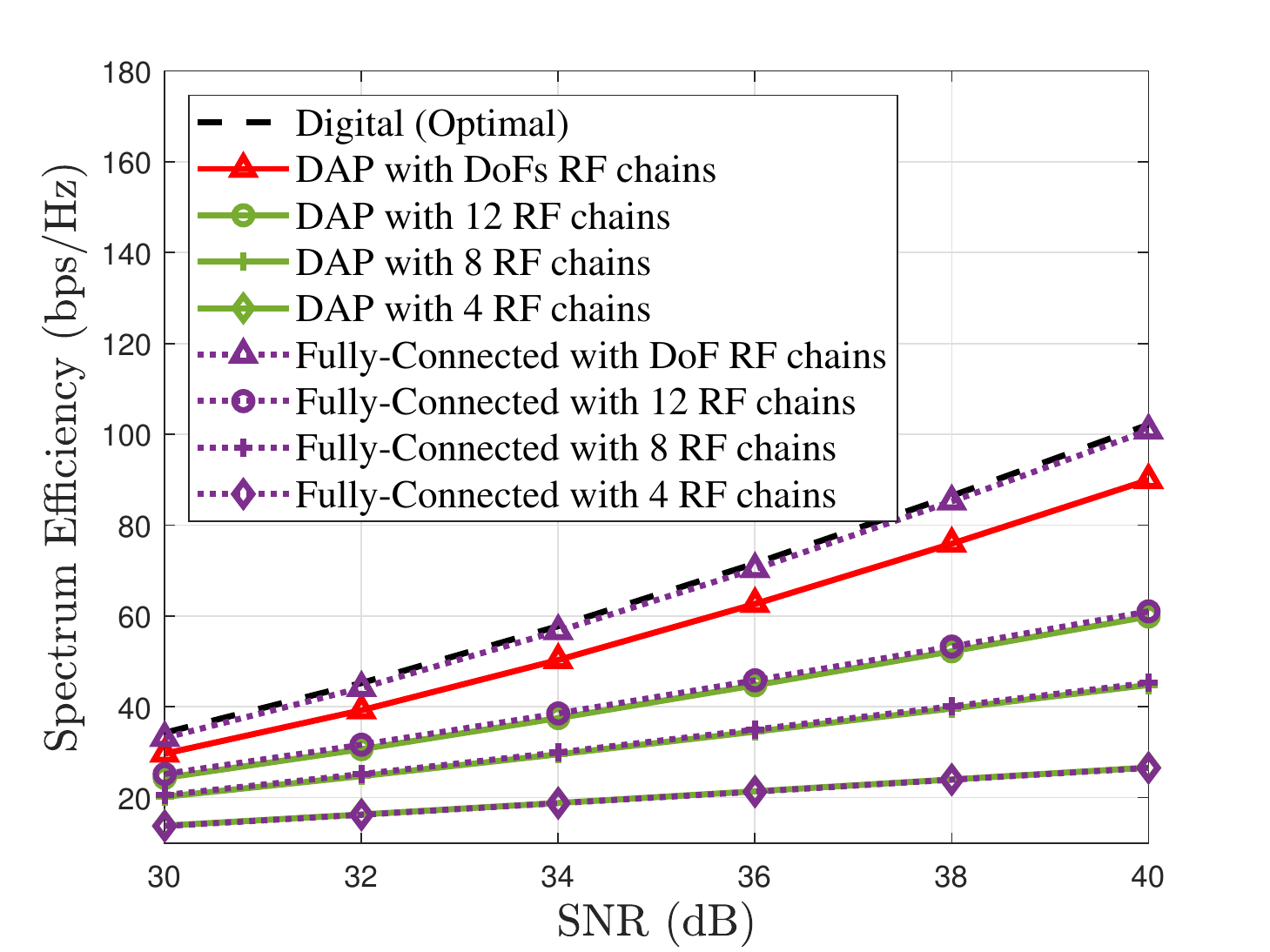}
	}
	\hspace{1in}
	\subfigure[Energy efficiency]{
		\label{img:ns_ee}
		\includegraphics[width=3in]{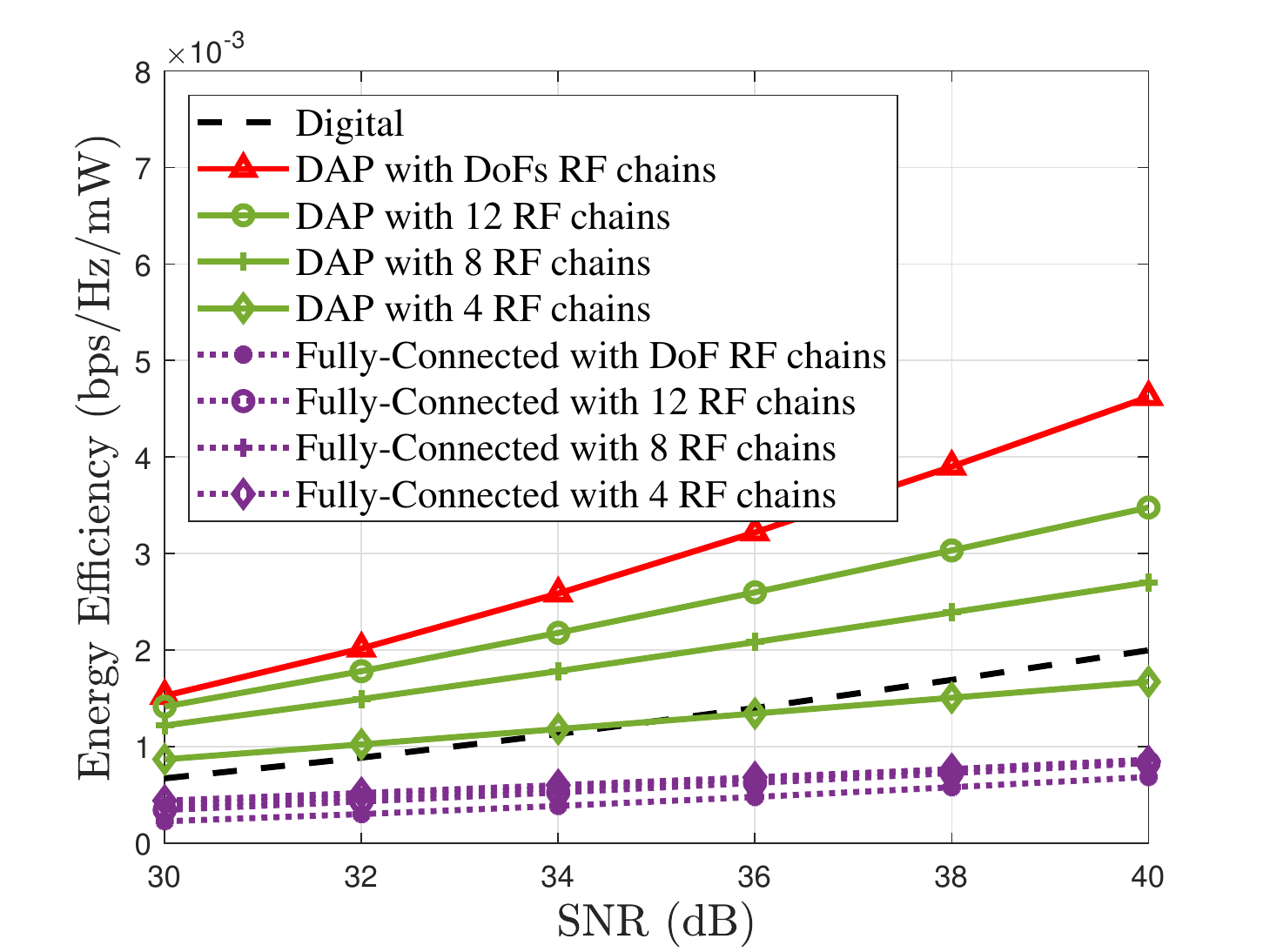}
	}
	\caption{The comparison of different $N_{\rm{s}}$ over spectrum efficiency and energy efficiency. The number of RF chains is chosen within $N_{\rm{DoF}}$, 12, 8, 4. The precoding architectures considered include fully-digital precoding architecture, proposed DAP architecture, fully-connected precoding architecture. The SNR varies from 30 dB to 40 dB.}
	\label{fig:subfig}
\end{figure}

 In the proposed DAP algorithm, we have assumed that $N_{\rm{s}}$ is equal to $N_{\rm{DoF}}$. To verify the rationality of choice, we compare different choices of $N_{\rm{s}}$ with $N_{\rm{s}} = N_{\rm{DoF}}, 12, 8, 4$. The transmission distance is $2$ m. The precoding baselines considered include the proposed DAP architecture and fully-connected precoding architecture with alternating precoding algorithm \cite{Yu'16'j}. As shown in Fig. \ref{img:ns_se}, the spectrum efficiency increases with increasing $N_{\rm{s}}$. With different $N_{\rm{s}}$, the performance of fully-connected precoding scheme outperforms our proposed sub-connected DAP architecture, which is consistent with previous works. When $N_{\rm{s}}$ is chosen to $N_{\rm{DoF}}$, the fully-connected precoding scheme approaches the optimal fully-digital precoding scheme, which reveals that the choice of $N_{\rm{s}}$ can be considered optimal. It is worth noting that, there is still a gap between the fully-connected precoding and proposed sub-connected DAP precoding scheme. However, as illustrated in Fig. \ref{img:ns_ee}, the proposed DAP architecture with $N_{\rm{s}} = N_{\rm{DoF}}$ outperforms all other settings in terms of energy efficiency. Therefore, the proposed DAP architecture can achieve spectrum efficiency and energy efficiency improvements with $N_{\rm{s}}$ equal to $N_{\rm{DoF}}$.
\section{Conclusions}\label{sec:conclusion}
In this paper, the significantly increased DoFs and the channel capacity in the near-field region are theoretically analyzed. To utilize the increased spatial DoFs, a DAP architecture and the corresponding precoding algorithm have been proposed for the XL-MIMO communication scenario, in which the number of activated RF chains can be flexibly adjusted to match the increased DoFs in the near-field region. Simulations results show that, compared with the classical sub-connected hybrid precoding architecture, the proposed DAP architecture can efficiently utilize the increased DoFs to significantly improve the spectrum efficiency in the near-field region. Different from alleviating problems brought by the near-field effect, this paper has revealed that the near-field property can be exploited to improve the performance of XL-MIMO communications, which may provide another methodology to meet the increasing demand of communication capacity. Our future research will focus on the capacity enhancement employing reconfigurable intelligent surface (RIS) considering the near-field effect.

\section*{Appendix A}
As shown in Fig. \ref{img:near-field}, according to the near-field channel in (\ref{eq: NF-channel}), the difference of the distance between $j$th and ($j+n$)th transmitter antenna with $i$th receiver antenna is written as 
\begin{equation}
\label{NF-distance}
\begin{aligned}
r_{pq} - r_{p(q+n)} &= r_{pq} \left( 1- \sqrt{1 + (\frac{nd}{r_{pq}})^2 - 2\frac{nd}{r_{pq}}\sin\phi_{pq}} \right) \\
                    &\mathop { \approx }\limits^{(a)} nd \sin\phi_{pq},
\end{aligned}
\end{equation}
where $\phi_{pq}$ denotes the azimuth angle between the $p^{th}$ receiver antenna and the $q$th transmitter antenna. Approximation (a) is derived based on the Taylor Series expansion that $\sqrt{1-x} \approx 1-\frac{1}{2}x $ when $x = 2\frac{nd}{r_{ij}}\sin\phi_{ij}$ is small enough. The approximation holds when $r_{ij} \gg nd$, which means that the size of the receiver antenna can be neglected compared with the communication distance. In the far-field range, the azimuth of different elements and the normalized path loss in the same antenna can be regarded as the same. Then, the $p^{th}$ row of $\bf{H}$ can be approximated as 
\begin{equation}
\label{channel approximation tx}
\begin{aligned}
{\bf{H}}_{p,:} = \gamma_{\rm{LoS}} \alpha e^{-j \frac{2\pi}{\lambda} r_{p1}} [1, e^{j \frac{2\pi}{\lambda}d\sin\phi_t}, \cdots, e^{j (N_t-1)\frac{2\pi}{\lambda}d\sin\phi_t}],
\end{aligned}
\end{equation}
where $\phi_{\rm{t}}$ and $\phi_{\rm{r}}$ represents the azimuth angle for all elements of the transmitter and receiver. The $q^{th}$ column of $\bf{H}$ can be formulated similarly. Thus, the near-field channel can be approximated as
\begin{equation}
\label{channel approximation rx}
\begin{aligned}
{\bf{H}}  = \gamma_{\rm{LoS}} \alpha e^{-j \frac{2\pi}{\lambda} r_{11}} {\bf{a}}_r(\phi_{\rm{r}}) {\bf{a}}_t(-\phi_{\rm{t}})^H.
\end{aligned}
\end{equation}
\par
Finally, by assuming $\alpha_{\rm{LoS}} = \alpha e^{-j \frac{2\pi}{\lambda} r_{11}}$ , $\phi_{\rm{LoS}}^{\rm{r}} = \phi_{\rm{r}}$ and $\phi_{\rm{LoS}}^{\rm{t}} = -\phi_{\rm{t}}$, the near-field LoS channel in (\ref{eq: NF-channel}) is degraded to LoS channel model in (\ref{eq: SV-channel}).

\section*{Appendix B}
\begin{figure}[!t]
	\centering
	\setlength{\abovecaptionskip}{0.cm}
	\includegraphics[width=3in]{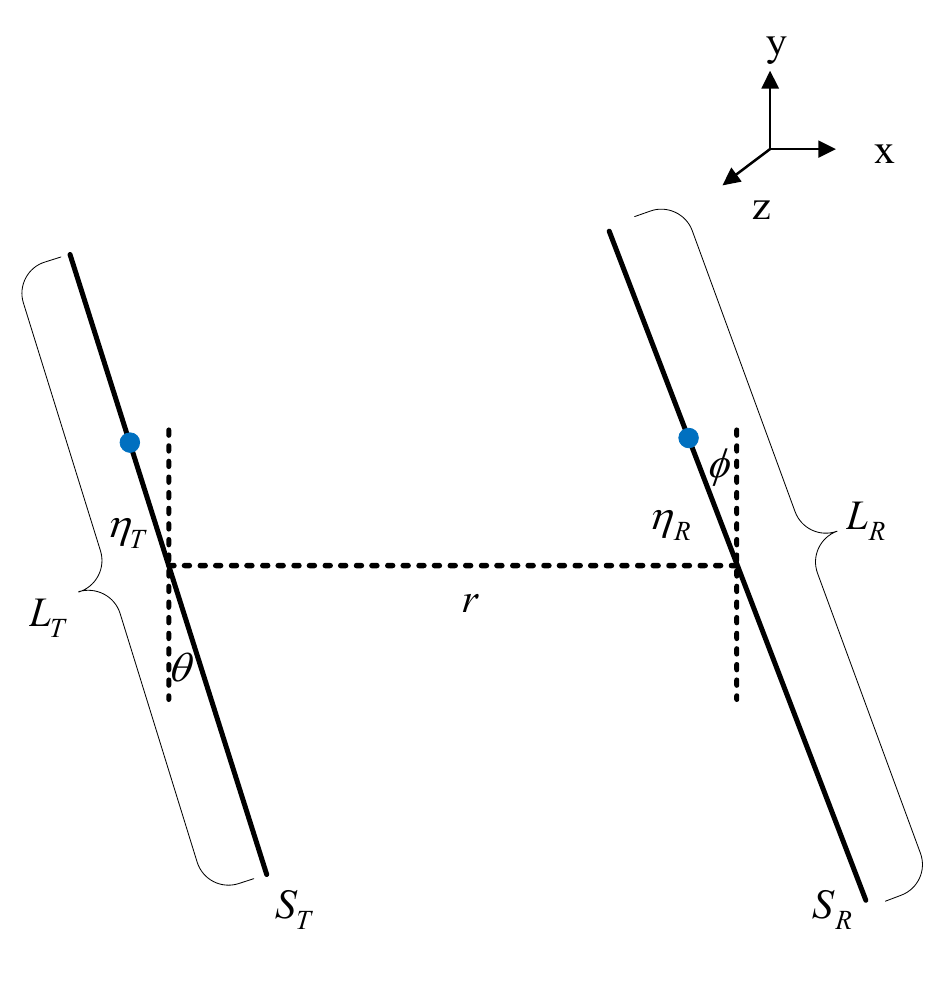}
	\caption{Non-parallel continuous linear arrays in the near-field region.}
	\label{img:array continuous}
\end{figure}
As shown in Fig. \ref{img:array continuous}, we consider a pair of non-parallel continuous arrays in which the center of both transmitter and receiver are in the same horizontal line. The length of the transmitter and receiver is $L_{\rm{t}}$ and $L_{\rm{r}}$ respectively. And the slant angle for transmitter and receiver is $\theta$ and $\phi$. Starting from the definition in (\ref{eq: definition of K 1}), we aim to maximize the power of the derived wave function
\begin{equation}
\label{eigenfunction}
\begin{aligned}
|g|^2 &= \int_{S_R} \phi^{*}({\bm{r}}_{\rm{R}})\phi({\bm{r}}_{\rm{R}})d{\bm{r}}_{\rm{R}} \\
& = \int_{S_R} \int_{S_T} \int_{S_T} G^{*}({\bm{r}}_{\rm{R}}, {\bm{r}}_{\rm{T}}')\psi^{*}({\bm{r}}_{\rm{T}}')G({\bm{r}}_{\rm{R}}, {\bm{r}}_{\rm{T}}) \\
& ~~~~\times \psi({\bm{r}}_{\rm{T}})d{\bm{r}}_{\rm{T}}'d{\bm{r}}_{\rm{T}}d{\bm{r}}_{\rm{R}}\\
& = \int_{S_T} \psi^{*}({\bm{r}}_T') \int_{S_T} K({\bm{r}}_T', {\bm{r}}_T) \psi({\bm{r}}_T) d{\bm{r}}_T'd{\bm{r}}_T.
\end{aligned}
\end{equation}
Then, to maximize the value of $|g|^2$, $\psi({\bm{r}}_T)$ has to be chosen as the eigenfunction corresponding to the largest eigenvalue \cite{miller'00'j}. The eigenfunction can be defined as 
\begin{equation}
\label{eq: definition of eigenfunction 2}
\begin{aligned}
|g|^2 \psi({\bm{r}}_T) &= \int_{S_T} K({\bm{r}}_T', {\bm{r}}_T) \psi({\bm{r}}_T') d{\bm{r}}_T' \\
& = \int_{S_T} \int_{S_R} \frac{{\rm{exp}}(jk|{\bm{r}}_R-{\bm{r}}_T|){\rm{exp}}(-jk|{\bm{r}}_R-{\bm{r}}_T'|)}{(4\pi)^2|{\bm{r}}_R-{\bm{r}}_T||{\bm{r}}_R-{\bm{r}}_T|} \\
&~~~~\times d{\bm{r}}_R \psi({\bm{r}}_T') d{\bm{r}}_T'.
\end{aligned}
\end{equation}
It is worth noting that, the green function $G({\bm{r}}_1, {\bm{r}}_2)$ is determined by the spherical wave assumption, which is also a basic assumption in the near-field communications. Then, since $|{\bm{r}}_R-{\bm{r}}_T||{\bm{r}}_R-{\bm{r}}_T|$ in the denominator is only a scalar factor, it can be viewed as $r^2$ if the transmission distance is relatively large. Then we concentrate on the nominator which contains a product of two phase items. Consider the variable $\eta_{\rm{t}}$ and $\eta_{\rm{r}}$ denote the distance from the center of transmitter and receiver respectively. Then we have
\begin{equation}
\label{eq: approx phase 1}
\begin{aligned}
& |{\bm{r}}_R-{\bm{r}}_T| = \\
& ~~\sqrt{(r + \eta_T \sin\theta - \eta_R \sin\phi)^2 + (\eta_T \cos \theta - \eta_R \cos \phi)^2}.
\end{aligned}
\end{equation}
Here we adopt the paraxial assumption that the distance between the two arrays are relatively large compared with their apertures. The first three items in Taylor series $\sqrt{1+x} \approx 1 + \frac{1}{2}x - \frac{1}{8}x^2$ are adopted. Then we have
\begin{equation}
\label{eq: approx phase 2}
\begin{aligned}
|{\bm{r}}_R-{\bm{r}}_T| &\approx r + \eta_T \sin\theta - \eta_R \sin\phi \\
&~~~~+ \frac{1}{2r}(\eta_T \cos \theta - \eta_R\cos\phi)^2.
\end{aligned}
\end{equation}
The product of the two phase items are formulated as
\begin{equation}
\label{eq: approx phase 3}
\begin{aligned}
&~~~\exp(jk|{\bm{r}}_R-{\bm{r}}_T|){\rm{exp}}(-jk|{\bm{r}}_R-{\bm{r}}_T'|) \\
&=\exp \big\{jk \big( (\eta_T - \eta_T')\sin \theta + \frac{\cos^2\theta}{2r}(\eta_T^2 - \eta_T^{'2}) \\
&~~~~- \frac{\cos\theta \cos\phi}{r}(\eta_T - \eta_T')\eta_R \big) \big\}.
\end{aligned}
\end{equation}
We assume a focusing function as
\begin{equation}
\label{eq: focusing function}
\begin{aligned}
F_T(\eta_T) = \exp \left\{jk \left(\eta_T \sin\theta + \frac{\cos^2\theta}{2r}\eta_T^2 \right) \right\}.
\end{aligned}
\end{equation}
Then we have
\begin{equation}
\label{eq: focusing function and others}
\begin{aligned}
\psi(\eta_T) = F_T(\eta_T)\alpha_T(\eta_T),
\end{aligned}
\end{equation}
where $\alpha_T(\eta_T)$ are the solution of the formulation below
\begin{equation}
\label{eq: alpha formulation}
\begin{aligned}
&|g|^2 (4\pi r)^2 \alpha_T(\eta_T) = \int_{-L_T/2}^{L_T/2} \int_{-L_R/2}^{L_R/2} \\
&~~~~ \exp \left\{jk \frac{\cos\theta\cos\phi}{r}(\eta_T - \eta_T')\eta_R \right\} d\eta_R \alpha_T(\eta_T') d\eta_T'.
\end{aligned}
\end{equation}
It can be easily derived that the integral
\begin{equation}
\label{eq: integral of eigenfunction}
\begin{aligned}
&~~~\int_{-L_R/2}^{L_R/2} {\rm{exp}} \left\{jk \frac{\cos\theta\cos\phi}{r}(\eta_T - \eta_T')\eta_R \right\} d\eta_R \\
& = \frac{\lambda r}{\cos\theta\cos\phi} \cdot \frac{\sin \Omega_T (\eta_T - \eta_T')}{\pi (\eta_T - \eta_T')}, 
\end{aligned}
\end{equation}
where $\Omega_T = \frac{k L_R\cos\theta\cos\phi}{2r}$. Thus, the eigenfunction can be rewritten as
\begin{equation}
\label{eq: final eigenfunction}
\begin{aligned}
|g|^2 (4\pi r)^2 \alpha_T(\eta_T) &= \int_{-L_T/2}^{L_T/2} \frac{\lambda r}{\cos\theta\cos\phi} \\
&~~~~\times \frac{\sin \Omega_T (\eta_T - \eta_T')}{\pi (\eta_T - \eta_T')}\alpha_T(\eta_T') d\eta_T'.
\end{aligned}
\end{equation}
Compared with the definition of prolate spheroidal wave functions in (\ref{eigenfunction prolate spheroidal}), parameter $c_y = \Omega_T \frac{L_T}{2} = \frac{\pi L_TL_R\cos\theta\cos\phi}{2\lambda r}$ is proportional to $\cos\theta\cos\phi$. Since $N_{\rm{DoF}}$ can be approximated as $N_{\rm{DoF}} = \frac{2}{\pi} c_y$. Thus, the formulation in (\ref{eq:critical value for rotation}) can be obtained. Noting that the derivative process is under paraxial assumptions. When the arrays are non-parallel, the paraxial degeneracy will be removed, resulting different eigenvalues $\upsilon_i$ in (\ref{eigenfunction prolate spheroidal}) when $i \leq \frac{2}{\pi} c_y$. However, since the eigenvalues $\upsilon_i$ fall off rapidly in an exponential manner for $i > \frac{2}{\pi} c_y$, the DoFs still remain unchanged for $\theta$ and $\phi$ far smaller than $\pi/2$.
\section*{Appendix C}
For a diagonal matrix ${\bf{F}}_{\rm{A}}$ and permuation matrix ${\bf{P}}_{\rm{S}}$, since it has been proved that for any matrix ${\bf{A}}$, ${\bf{B}} = {\bf{P}}_{\rm{S}}^{T} {\bf{A}} {\bf{P}}_{\rm{S}}$ has the same diagonal entities as ${\bf{A}}$ (but may with different orders) \cite{Brewer'78}. Thus, for any diagonal matrix ${\bf{F}}_{\rm{A}}$, there exists a diagonal matrix $\widetilde{\bf{F}}_{\rm{A}}$ that has the same entities with ${\bf{F}}_{\rm{A}}$ subjecting to
\begin{equation*}
\label{eq: permutation proof}
\begin{aligned}
{\bf{F}}_{\rm{A}} {\bf{P}}_{\rm{S}} = &~~{\bf{P}}_{\rm{S}} \left[{\bf{P}}_{\rm{S}}^{T} {\bf{F}}_{\rm{A}} {\bf{P}}_{\rm{S}} \right] = {\bf{P}}_{\rm{S}} \widetilde{\bf{F}}_{\rm{A}}.
\end{aligned}
\end{equation*}
Therefore, the order of the permuation matrix ${\bf{P}}_{\rm{S}}$ and diagonal analog precoder ${\bf{F}}_{\rm{A}}$ can be switched for further processings. 

\footnotesize
\balance 
\bibliographystyle{IEEEtran}
\bibliography{IEEEabrv,reference}

\begin{thebibliography}{10}
\providecommand{\url}[1]{#1}
\csname url@samestyle\endcsname
\providecommand{\newblock}{\relax}
\providecommand{\bibinfo}[2]{#2}
\providecommand{\BIBentrySTDinterwordspacing}{\spaceskip=0pt\relax}
\providecommand{\BIBentryALTinterwordstretchfactor}{4}
\providecommand{\BIBentryALTinterwordspacing}{\spaceskip=\fontdimen2\font plus
\BIBentryALTinterwordstretchfactor\fontdimen3\font minus
  \fontdimen4\font\relax}
\providecommand{\BIBforeignlanguage}[2]{{%
\expandafter\ifx\csname l@#1\endcsname\relax
\typeout{** WARNING: IEEEtran.bst: No hyphenation pattern has been}%
\typeout{** loaded for the language `#1'. Using the pattern for}%
\typeout{** the default language instead.}%
\else
\language=\csname l@#1\endcsname
\fi
#2}}
\providecommand{\BIBdecl}{\relax}
\BIBdecl

\bibitem{sun'14'j}
S.~{Sun}, T.~S. {Rappaport}, R.~W. {Heath}, A.~{Nix}, and S.~{Rangan}, ``{MIMO}
  for millimeter-wave wireless communications: beamforming, spatial
  multiplexing, or both?'' \emph{{IEEE} Commun. Mag.}, vol.~52, no.~12, pp.
  110--121, Dec. 2014.

\bibitem{bjornson'2019'j}
E.~{Bj{\"o}rnson}, L.~{Sanguinetti}, H.~{Wymeersch}, J.~{Hoydis}, and T.~L.
  {Marzetta}, ``Massive {MIMO} is a reality—what is next?: Five promising
  research directions for antenna arrays,'' \emph{Digit. Signal Process.},
  vol.~94, pp. 3--20, Nov. 2019.

\bibitem{Ayach'14'j}
O.~{Ayach}, S.~{Rajagopal}, S.~{Abu-Surra}, Z.~{Pi}, and R.~W. {Heath},
  ``Spatially sparse precoding in millimeter wave {MIMO} systems,''
  \emph{{IEEE} Trans. Wireless Commun.}, vol.~13, no.~3, pp. 1499--1513, Jan.
  2014.

\bibitem{Sherman'62'j}
J.~{Sherman}, ``Properties of focused apertures in the fresnel region,''
  \emph{{IEEE} Trans. Antennas Propag.}, vol.~10, no.~4, pp. 399--408, Jul.
  1962.

\bibitem{cui'21}
M.~Cui, L.~Dai, R.~Schober, and L.~Hanzo, ``Near-field wideband beamforming for
  extremely large antenna array,'' \emph{arXiv preprint arXiv:2109.10054}, Sep.
  2021.

\bibitem{headland'18't}
D.~{Headland}, Y.~{Monnai}, D.~{Abbott}, C.~{Fumeaux}, and
  W.~{Withayachumnankul}, ``Tutorial: Terahertz beamforming, from concepts to
  realizations,'' \emph{APL Photon.}, vol.~3, no.~5, 2018, Art. No. 051101.

\bibitem{zhang'21'}
H.~{Zhang}, N.~{Shlezinger}, F.~{Guidi}, D.~{Dardari}, M.~F. {Imani}, and Y.~C.
  {Eldar}, ``Beam focusing for near-field multi-user {MIMO} communications,''
  \emph{arXiv preprint arXiv:2105.13087}, May 2021.

\bibitem{cui'22'j}
M.~{Cui} and L.~{Dai}, ``Channel estimation for extremely large-scale {MIMO}:
  Far-field or near-field?'' \emph{IEEE Trans. Commun. (early access)}, pp.
  1--1, Jan. 2022.

\bibitem{Yu'20'j}
Y.~{Han}, S.~{Jin}, C.~{Wen}, and X.~{Ma}, ``Channel estimation for extremely
  large-scale massive {MIMO} systems,'' \emph{{IEEE} Commun. Lett.}, vol.~9,
  no.~5, pp. 633--637, May 2020.

\bibitem{Wei'21'j}
X.~{Wei} and L.~{Dai}, ``Channel estimation for extremely large-scale massive
  {MIMO}: Far-field, near-field, or hybrid-field?'' \emph{{IEEE} Commun.
  Lett.}, vol.~26, no.~1, pp. 177--181, Jan. 2022.

\bibitem{han'20'j}
Y.~{Han}, S.~{Jin}, C.~{Wen}, and X.~{Ma}, ``Channel estimation for extremely
  large-scale massive {MIMO} systems,'' \emph{{IEEE} Commun. Lett.}, vol.~9,
  no.~5, pp. 633--637, May 2020.

\bibitem{zhangzz'21'}
Z.~{Zhang} and L.~{Dai}, ``Continuous-aperture {MIMO} for electromagnetic
  information theory,'' \emph{arXiv preprint arXiv:2111.08630}, Nov. 2021.

\bibitem{Davide'20'jsac}
D.~{Davide}, ``Communicating with large intelligent surfaces: Fundamental
  limits and models,'' \emph{{IEEE} J. Sel. Areas Commun.}, vol.~38, no.~11,
  pp. 2526--2537, Nov. 2020.

\bibitem{miller'00'j}
D.~A.~B. {Miller}, ``Communicating with waves between volumes: evaluating
  orthogonal spatial channels and limits on coupling strengths,'' \emph{Appl.
  Opt.}, vol.~39, no.~11, pp. 1681--1699, Apr. 2000.

\bibitem{Decarli'21}
N.~Decarli and D.~Dardari, ``Communication modes with large intelligent
  surfaces in the near field,'' \emph{arXiv preprint arXiv:2108.10569}, Aug.
  2021.

\bibitem{Heath'16'j}
R.~W. {Heath}, N.~{González-Prelcic}, S.~{Rangan}, W.~{Roh}, and A.~M.
  {Sayeed}, ``An overview of signal processing techniques for millimeter wave
  {MIMO} systems,'' \emph{{IEEE} J. Sel. Topics Signal Process.}, vol.~10,
  no.~3, pp. 436--453, Apr. 2016.

\bibitem{Slepian'1961'j}
D.~{Slepian} and H.~O. {Pollak}, ``Prolate spheroidal wave functions, fourier
  analysis and uncertainty — {I},'' \emph{The Bell System Technical Journal},
  vol.~40, no.~1, pp. 43--63, Jan. 1961.

\bibitem{marzetta'2016'fundamentals}
T.~L. {Marzetta} and H.~Q. {Ngo}, \emph{Fundamentals of massive MIMO}.\hskip
  1em plus 0.5em minus 0.4em\relax Cambridge University Press, 2016.

\bibitem{miller'19'waves}
D.~A. {Miller}, ``Waves, modes, communications, and optics: a tutorial,''
  \emph{Adv. in Opt. and Photon.}, vol.~11, no.~3, pp. 679--825, Sep. 2019.

\bibitem{Gao'16'j}
X.~{Gao}, L.~{Dai}, S.~{Han}, {C.-L. I}, and R.~W. {Heath}, ``Energy-efficient
  hybrid analog and digital precoding for mmwave {MIMO} systems with large
  antenna arrays,'' \emph{{IEEE} J. Sel. Areas Commun.}, vol.~34, no.~4, pp.
  998--1009, Apr. 2016.

\bibitem{wolkowicz'80'}
H.~{Wolkowicz} and G.~P. {Styan}, ``Bounds for eigenvalues using traces,''
  \emph{Linear Algebra Appl.}, vol.~29, pp. 471--506, Feb. 1980.

\bibitem{Sun'18'c}
Y.~{Sun}, Z.~{Gao}, H.~{Wang}, and D.~{Wu}, ``Machine learning based hybrid
  precoding for mmwave {MIMO-OFDM} with dynamic subarray,'' in \emph{Proc. IEEE
  Globecom Workshops (GC Wkshps)}, Dec. 2018, pp. 1--6.

\bibitem{park'17'j}
S.~{Park}, A.~{Alkhateeb}, and {R. W. Heath}, ``Dynamic subarrays for hybrid
  precoding in wideband mmwave {MIMO} systems,'' \emph{{IEEE} Trans. Wireless
  Commun.}, vol.~16, no.~5, pp. 2907--2920, May 2017.

\bibitem{Xu'19'c}
K.~{Xu}, F.~C. {Zheng}, P.~{Cao}, H.~{Xu}, and X.~{Zhu}, ``A low complexity
  greedy algorithm for dynamic subarrays in mmwave {MIMO} systems,'' in
  \emph{Proc. IEEE 90th Veh. Technol. Conf. (IEEE VTC'19)}, Sep. 2019, pp.
  1--5.

\bibitem{Yu'16'j}
X.~{Yu}, J.~Z. J.~{Shen}, and K.~B. {Letaief}, ``Alternating minimization
  algorithms for hybrid precoding in millimeter wave {MIMO} systems,''
  \emph{{IEEE} J. Sel. Areas Commun.}, vol.~10, no.~3, pp. 485--500, Apr. 2016.

\bibitem{Brewer'78}
J.~{Brewer}, ``Kronecker products and matrix calculus in system theory,''
  \emph{{IEEE} Trans. Circuits Syst.}, vol.~25, no.~9, pp. 772--781, Sep. 1978.

\end{thebibliography}
\end{document}